\documentstyle[preprint,aps,epsfig]{revtex}
\begin{document}
\draft

\title{Slow Drag in 2D Granular Media}

\author{Junfei Geng and R.~P. Behringer}

\address{Department of Physics and Center for Nonlinear and Complex Systems, Duke University, Durham NC, 27708-0305, USA}

\date{\today}
\maketitle

\begin{abstract}

We study the drag force experienced by an object slowly moving at
constant velocity through a 2D granular material consisting of
bidisperse disks. The drag force is dominated by force chain
structures in the bulk of the system, thus showing strong
fluctuations. We consider the effect of three important control
parameters for the system: the packing fraction, the drag velocity and
the size of the tracer particle. We find that the mean drag force
increases as a power-law (exponent of 1.5) in the reduced packing
fraction, $(\gamma - \gamma_c)/\gamma_c$, as $\gamma$ passes through a
critical packing fraction, $\gamma_c$. By comparison, the mean drag
grows slowly (basically logarithmic) with the drag velocity, showing a
weak rate-dependence. We also find that the mean drag force depends
nonlinearly on the diameter, $a$ of the tracer particle when $a$ is
comparable to the surrounding particles' size. However, the system
nevertheless exhibits strong statistical invariance in the sense that
many physical quantities collapse onto a single curve under
appropriate scaling: force distributions P($f$) collapse with
appropriate scaling by the mean force, the power spectra P($\omega$)
collapse when scaled by the drag velocity, and the avalanche size and
duration distributions collapse when scaled by the mean avalanche size
and duration. We also show that the system can be understood using
simple failure models, which reproduce many experimental
observations. These observations include: a power law variation of the
spectrum with frequency characterized by an exponent $\alpha=-2$,
exponential distributions for both the avalanche size and duration,
and an exponential fall-off at large forces for the force
distributions. These experimental data and simulations indicate that
fluctuations in the drag force seem to be associated with the force
chain formation and breaking in the system. Moreover, our simulations
suggest that the logarithmic increase of the mean drag force with rate
can be accounted for if slow relaxation of the force chain networks is
included.

\end{abstract}

\pacs{PACS numbers: 46.10.+z, 47.20.-k}

\section{introduction}

Granular materials are of great interest for their rich phenomenology
and import applications \cite{reviews}. When subject to external
stresses, a dense granular system forms inhomogeneous force chain
networks where only a fraction of the grains carry most of the force
\cite{chains}. The spatial scale of these force chains can extend over
many grain diameters, and the chain lengths may be comparable to the
system size. The separation between microscopic and macroscopic scales
poses a theoretical challenge if one attempts to describe a granular
system using continuum approach. Recently, experimental works by
several research groups
\cite{fluct_expts,mueth_98,miller_96,howell_99,hartley_03,albert_00}
suggest the importance of strong stress fluctuations in granular
systems. The fluctuations, as characterized by the standard deviation
or $rms$ of the stress, can often be somewhere from 1 to several times
of the mean stress. However, questions involving the dynamics, nature,
and length/time scales associated with these fluctuations are still
poorly understood. An improved understanding of these questions could
provide insight into describing a number of practical applications and
such phenomena as earthquakes and avalanches. Another motivation
concerns exploring jamming \cite{liu_98,cates_98} in granular
materials. Specifically, jammed states in granular systems may be
reached when the density (packing fraction) of the system is high
enough.

In this regard, slow drag experiments, the subject of this paper,
provide a useful way to understand the nature of stress fluctuations
and slow dynamics in granular materials. We have used a similar
experimental approach to probe the thermodynamic temperature in
granular systems, as reported elsewhere \cite{geng_03}.

In molecular fluids, the drag force on a particle arises from viscous
interactions, i.e. from collisional interactions of the particle and
surrounding molecules that involve momentum transfer.  This drag force
is linearly proportional to the object's velocity through the fluid
when the velocity is not very large.  

In dense granular media, the origin of the drag force differs in
several respects.  First, frictional interactions exist between a drag
particle and surrounding grains.  Second, but related, is the
existence of force chains.  These relatively long-range inhomogeneous
structures can provide an elastic (rigid in the limit of infinitely
stiff particles) resistance to a moving particle.

In Fig. \ref{fig:setup}c, we show such force chain structures obtained
using photoelastic techniques \cite{howell_99,geng_01}. These force
chains are typically inhomogeneous and anisotropic in nature, and
constantly form and break when an object moves through the granular
media, leading to strong fluctuations in the drag force. In the
experiments presented here, we consider the drag force experienced by
a tracer particle moving through a 2D granular material consisting of
bidisperse disks.  In our experiment, the size of the tracer particle
is comparable to the surrounding grains, which allows us to explore
fluctuations at the grain scale. The experimental results presented
here are described well by simple failure models.

A number of experimental and theoretical results provide important
background to the present studies.  Experiments that are relevant here
include the ``carbon paper'' studies of Mueth et al. \cite{mueth_98},
who measured the static forces of a material (e.g. glass beads) at the
boundary of a container, and showed that the distribution of forces,
$f$, is exponential for large $f$. Sheared granular systems, both in
2D \cite{howell_99,hartley_03} and 3D \cite{miller_96}, show strong
force/stress fluctuations. In addition, the 2D experiments by Howell
et al. \cite{howell_99} showed a well defined strengthening/softening
transition as the packing fraction of the system passed a critical
packing fraction $\gamma_c$. The mean stress in such a system varies
as a power-law in the reduced packing fraction,
\begin{equation}
r=\frac{\gamma-\gamma_c}{\gamma_c},
\end{equation}
with an exponent between 2 and 4, depending on the particle
type. Later experiments on similar 2D systems by Hartley et
al. \cite{hartley_03} showed that the mean stress increased
logarithmically with the shearing rate, which may be related to
collective slow relaxation of the force chain network.  3D experiments
by Miller et al. \cite{miller_96} identified rate-independent power
spectra, $P(\omega)$, for the stress time series which fell off as $P
\sim \omega^{-2}$ at high spectral frequency $\omega$. Experiments on
3D drag by Albert et al. \cite{albert_99,albert_00,albert_01} relate
most closely to the present experiments. These studies yielded the
drag force experienced by a rod as it was dragged through granular
materials such as glass beads. Depending on the rod insertion depth
and the size ratio between the rod and the grain, three types of drag
force time series were observed: a periodic regime where the signal
resembles an ideal sawtooth pattern, a random regime, and a stepped
regime with sawtooth-like steps. These authors focused their work on
the periodic and stepped regimes, characterized by stick-slip
fluctuations due to successive formation and collapse of jammed
states. A particularly interesting finding of these studies was that
the mean drag force on the rod was independent of the drag velocity.

Several theoretical works \cite{q-model,kahng_01} have provided a
context for understanding the stress distributions and stress
fluctuations in granular materials. The q-model of Coppersmith et
al. \cite{q-model} predicts a force distribution for static systems
$P(F)\propto F^{N-1}$exp$(-F/F_0)$, where $N$ is the system
dimension. This model only considers the vertical force transmitted
through a regularly packed lattice. Vertical forces on a grain in one
layer are balanced by transmitting fractions, $q$ and (1-$q$), to the
two supporting grains in the next layer (assuming a 2D system), where
$q$ is a random number uniformly distributed in $0\le q\le 1$. We note
for exponential force distributions, that the mean is of the order of
the width of the distribution.  Other lattice models
\cite{other_lattice} and calculations by Radjai using contact dynamics
\cite{radjai_96} also predict exponential force distributions for
large forces.

Recently, Kahng et al. \cite{kahng_01} have used a stochastic failure
model to understand the 3D drag experiments of Albert et
al. \cite{albert_99,albert_00,albert_01}. These authors used simple
springs with random thresholds to model the jamming and reorganization
of grains. Among other results, the model reproduces the
experimentally observed periodic sawtooth fluctuations in the drag
force. We will use this simple failure model, with modifications,
later in this paper to understand the experiments described here.

The organization of the remainder of this paper is as follows. In
Section II, we describe the experimental setup and procedures. In
Section III, we report experimental results. In Section IV, we
describe models and simulations. Finally, we draw conclusions in
Section V.

\section{Experimental setup and procedures}

\begin{figure}[h]
\center{\parbox{5in}{\psfig{file=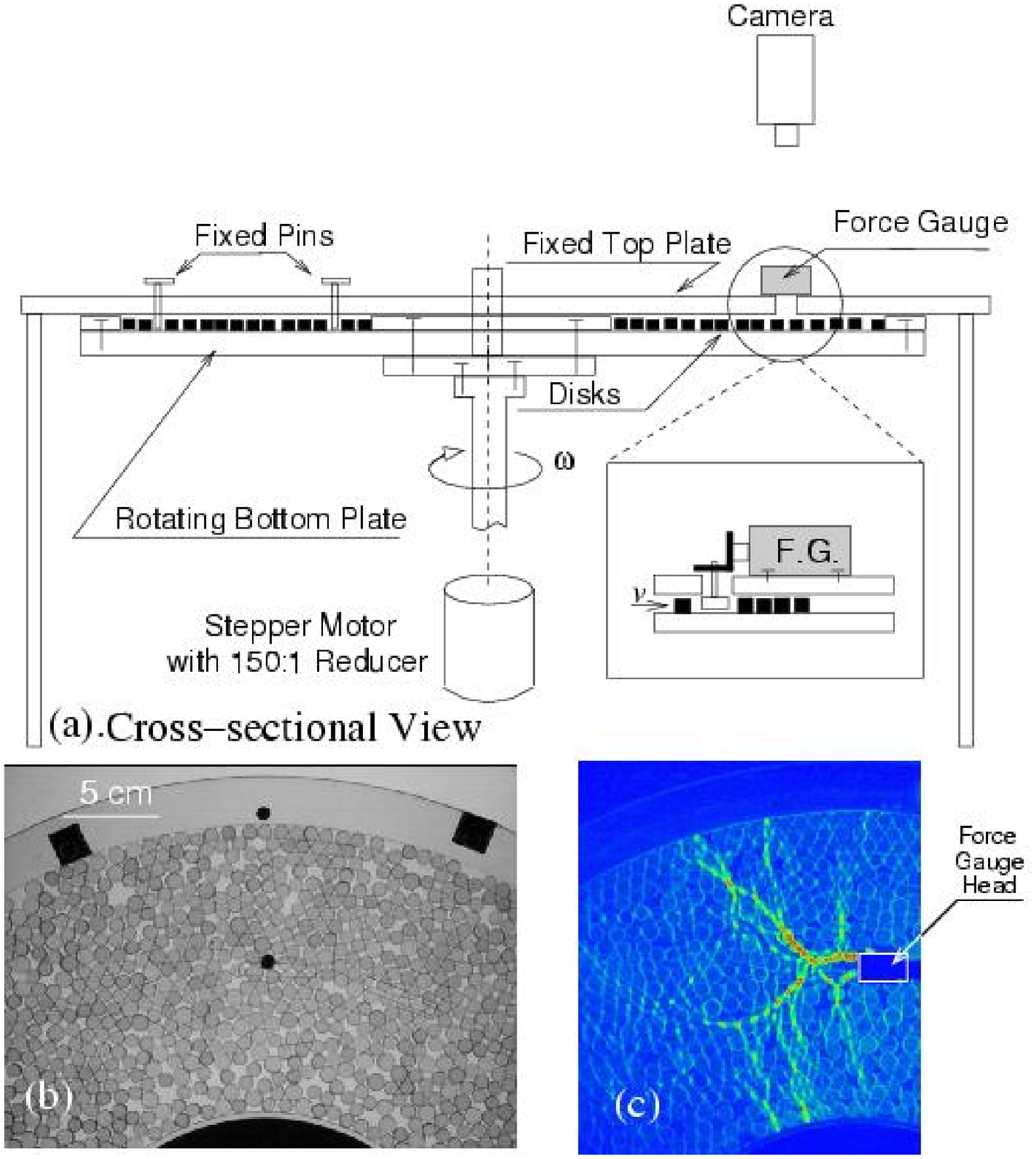,width=5in}}}
\caption{(a). Schematic drawings of the apparatus: a cross-sectional
view, where the plane of the section is through a diameter of the
apparatus, which has circular symmetry in the horizontal plane. The
bottom plate, together with particles, rotate as a rigid body at a
slow velocity. The inset shows how a digital force gauge (F.G.) is
mounted on the top plate and connected with the tracer particle
through the force gauge hole. (b). An actual image taken from the
experiments showing the 2D granular system composed of bidisperse
disks. (c). A stress image, obtained using
photoelasticity\protect\cite{howell_99,geng_01}, showing force chain
structures when the tracer particle is dragged through the medium.}
\label{fig:setup}
\end{figure}

The experiments were carried out in an apparatus which is, in spirit,
similar to the one in Ref.~\cite{albert_99}, except that the one used
here is two-dimensional in character, whereas the one used by Albert
et al. was three-dimensional. We show a cross-sectional view of the
apparatus in Fig.~\ref{fig:setup}a. The bottom plate was driven by the
center shaft, both of which are supported by ball bearings mounted on
a stable metal table (not shown). A stepper motor ran at a low
frequency to drive the bottom plate. The top plate did not rotate and
had no contact with the rotating bottom plate or the particles. The
granular medium consisted of a single layer of bi-disperse disks with
diameters 0.744 and 0.876 cm, where the thickness of both types of
disk was 0.660 cm. Fig.~\ref{fig:setup}b shows an actual image from
the experiment where the two types of disks can be identified. The
disks were placed on the bottom plate and confined between two
concentric ring structures. The inner ring radius was 10.5 cm and the
outer ring radius was 25.4 cm. When the bottom plate was rotating, the
disks moved with it as a rigid body, due to friction.  This frictional
force with the substrate was relatively weak compared to the forces
between particles associated with force chains. The centrifugal force
experienced by the disks was negligible due to the slow rotation
speed. Note that this apparatus is not to be confused with a Couette
shearing apparatus where either the inner wheel or the outer wheel is
moving. In this apparatus, both inner and outer boundaries remained
fixed and the driving was provided by the moving bottom plate.

A digital force gauge (Model DPS-110 from Imada Inc., resolution 0.1
g), shown in the inset of the Fig.~\ref{fig:setup}a, was mounted on
one side of the top plate. The force sensor was connected with a
tracer particle through a hole, located in the center of the inner and
outer ring. The reading on the force gauge, which yielded the
instantaneous tangential force, was recorded as a time series by a
computer through its serial communication port, as in
Fig. \ref{fig:force_series}. When the granular medium moved, force
chains form in the bulk of the system, as show in
Fig.~\ref{fig:setup}c using photoelastic
techniques~\cite{howell_99,geng_01}.  The pins on the left side of the
top plate stirred the particles.

There are three important parameters that we explored in the system,
i.e, the rotation rate $\omega$, the system packing fraction $\gamma$
(or density), and the tracer particle size $a$. We varied the rotation
rate $\omega$ over two orders of magnitude, from $\omega=6.33\times
10^{-6}$ to $8.67\times 10^{-4}Hz$ (corresponding to $v=7.14 \times
10^{-6}$ to $ 9.78 \times 10^{-4}$ $m/s$), the packing fraction
$\gamma$ from 0.561 to 0.761 (these values are global packing
fractions since the system is not completely uniform), and the tracer
particle diameters over the set of diameters $a=0.744$, $0.876$,
$1.250$, $1.610$ and $1.930$ cm.

\section{Experimental results}

In this section, we report the experimental results. We first consider
the effect of rotation rate, and we then turn to the effect of changes
in the packing fraction.

\subsection{Changing the Medium Rotation Rate}

\begin{figure}[h]
\center{\parbox{5in}{\psfig{file=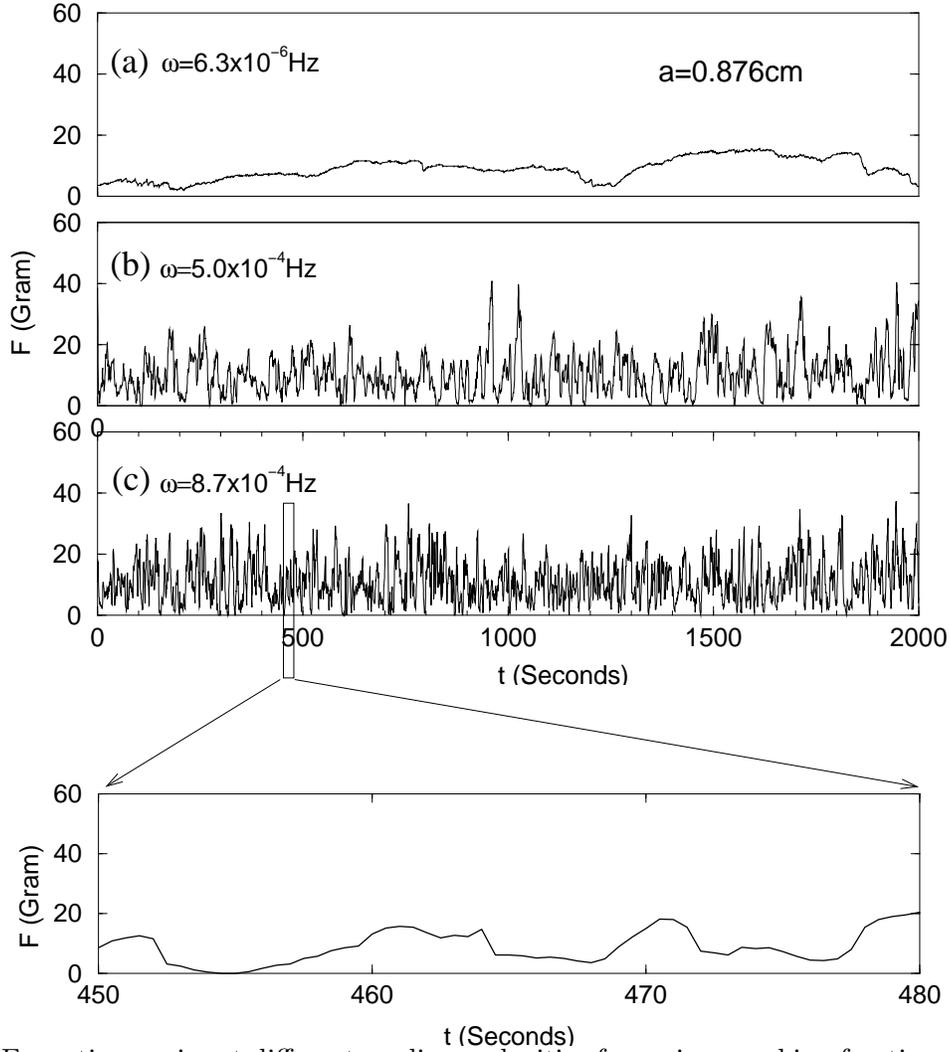,width=5in}}}
\caption[Force time series at different medium rotation rates]{Force
time series at different medium velocities for a given packing
fraction ($\gamma=0.754$) and a given tracer size ($a=0.876$ cm). The
rotation rates are: (a). $\omega=6.3\times 10^{-6} Hz$,
(b). $\omega=5.0\times 10^{-4}Hz$, and (c). $\omega=8.7\times
10^{-4}Hz$. The velocity is obtained according to $v=\omega r$, where
$r=17.95$ cm. These force series show strong fluctuations.  An
enlarged view of a small segment of (c) also suggests self-similar
structures in the system.}
\label{fig:force_series}
\end{figure}

An initial series of experiments was carried out at a fixed packing
fraction $\gamma=0.754$, which is above the critical packing fraction
$\gamma_c$, discussed in more detail in the next section.  Here, we
varied the rotation rate over $\omega=6.33\times 10^{-6} \leq \omega
\leq 8.67\times 10^{-4}Hz$ (corresponding to $7.14 \times 10^{-6} \leq
v \leq 9.78 \times 10^{-4}$ $m/s$).  (The velocity of the tracer is
$v=\omega r$, where $r=17.95$ cm is the radial location of the
tracer.) In Fig. \ref{fig:force_series}, we show three sets of force
time series, obtained with a tracer size $a=0.876$ cm, and rotation
rates that spanned the full range of $\omega$'s, namely
(a). $\omega=6.3\times 10^{-6} Hz$, (b). $\omega=5.0\times 10^{-4}Hz$,
and (c). $\omega=8.7\times 10^{-4}Hz$.  As one would expect, the force
time series in Fig. \ref{fig:force_series} show strong
fluctuations. Interestingly, an enlarged view of a small section of
Fig. \ref{fig:force_series}c, seems qualitatively similar to the
slower run in Fig. \ref{fig:force_series}a, which suggests possible
scaling behavior.  We will return to this point below.

\subsubsection{Mean Drag Force and Force Distributions}

\begin{figure}[h]
\center{\parbox{4in}{\psfig{file=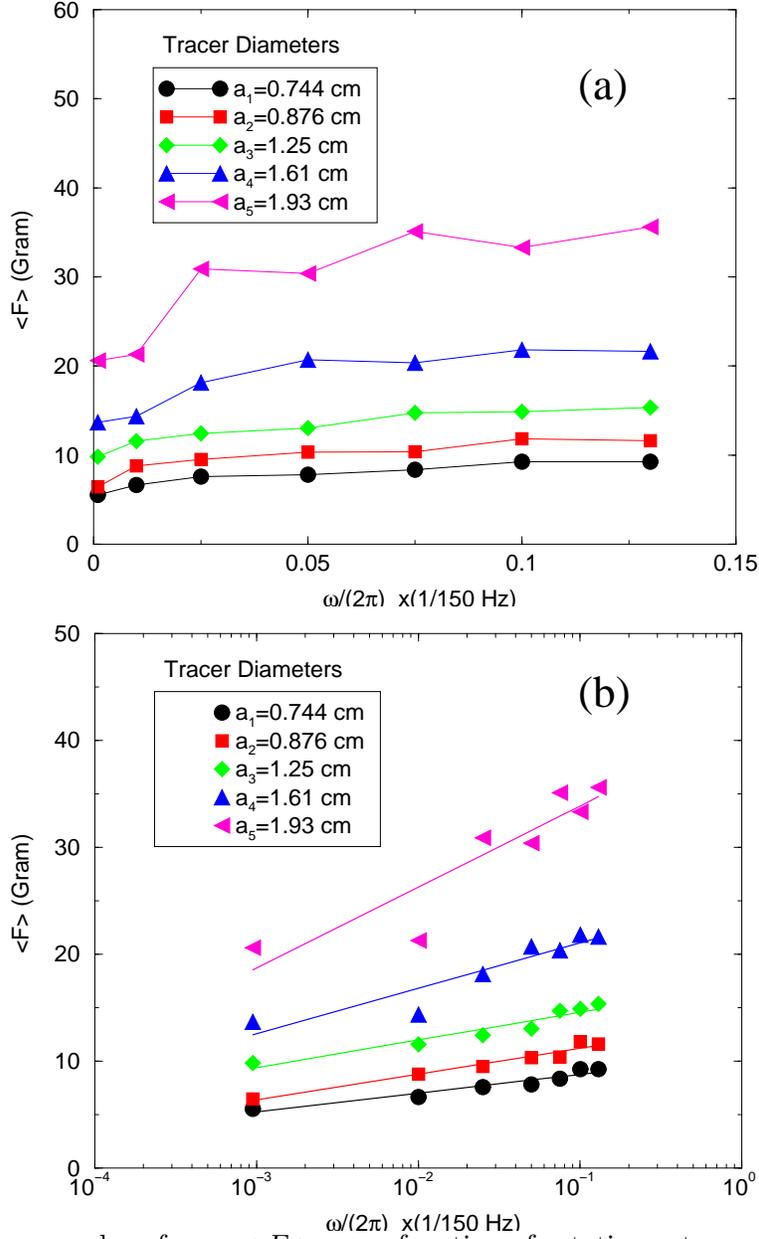,width=4in}}}
\caption{(a) The mean drag force, $<F>$, as a function of rotation
rate, $\omega$, for tracer particles with different sizes ($a=0.744$,
$0.876$, $1.25$, $1.61$ and $1.93$ cm.). (b) Same data as (a), but on
log-lin scales to emphasize that the mean force increases slowly
(basically logarithmically) with the medium velocity.  Throughout, we
use a forces normalized by g, the acceleration of gravity.}
\label{fig:f_v}
\end{figure}

In Fig. \ref{fig:f_v}a, we show the mean drag force, $<F>$, as a
function of rotation rate, $\omega$, for tracer particles of five
difference diameters ($a=0.744$, $0.876$, $1.25$, $1.61$ and $1.93$
cm.). For each of these tracer sizes, the mean drag force increased
only slightly (by a factor less than 2) for a variation by more than
two decades in $\omega$. To emphasize this slow increase, we plot the
same data on log-lin scales in Fig. \ref{fig:f_v}b. The data can be
fitted by a straight line, indicating a logarithmic variation of $<F>$
with $\omega$.  This is consistent with the results by Hartley et al.
\cite{hartley_03} who found that the total stress in a system of
similar particles undergoing slow shearing also increases
logarithmically with the shearing rate.

We emphasize that this slow increase in the mean force differs
significantly from the drag force in a fluid, where the mean force
increases linearly with the drag velocity when the velocity is not too
large. This is also in contrast to rate-independent stresses in
Mohr-Coulomb friction models \cite{nedderman,wood} for dense granular
systems.  It is consistent with several rate-dependent friction
models\cite{rate-dep}.

\begin{figure}[h]
\center{\parbox{4in}{\psfig{file=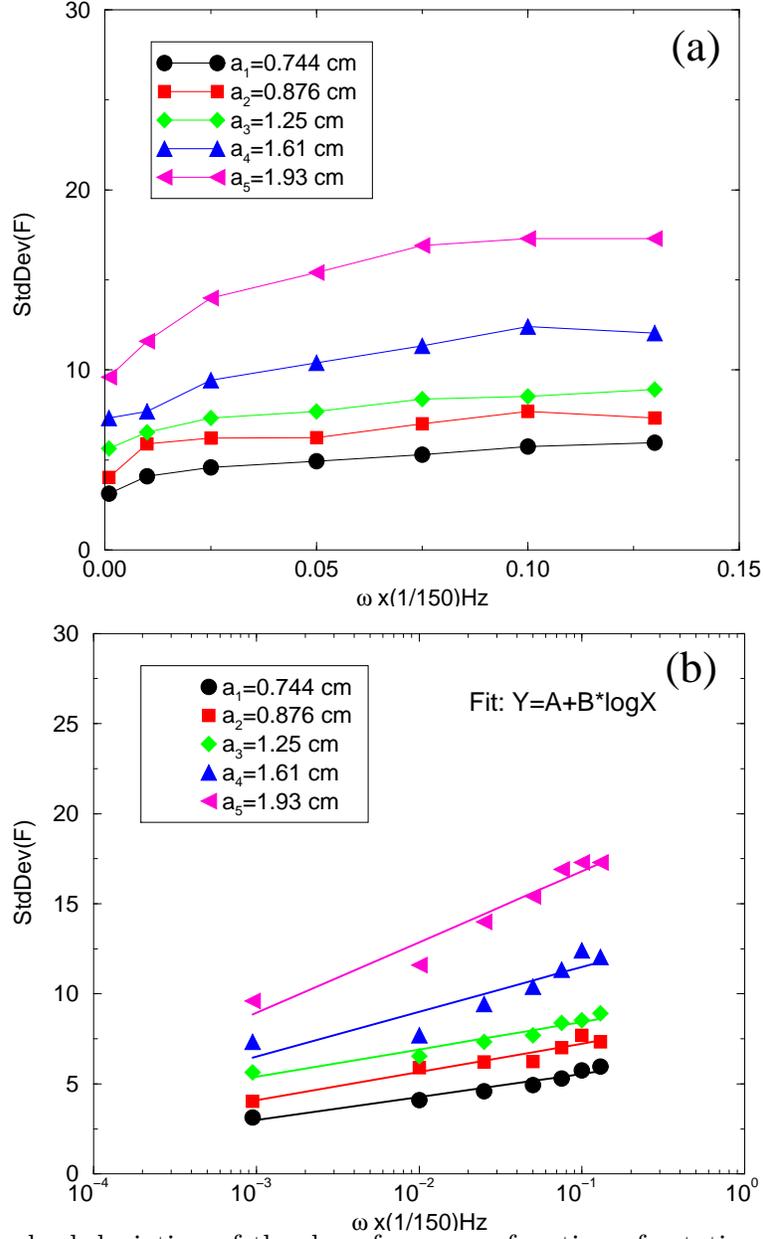,width=4in}}}
\caption{(a) Standard deviation of the drag force as a function of
rotation rate, $\omega$, for tracer particles with different sizes
($a=0.744$, $0.876$, $1.25$, $1.61$ and $1.93$ cm.). (b) Same data as
(a), but plotted on log-lin scales to emphasize that the standard
deviation also increases logarithmically with the medium velocity.}
\label{fig:variance_v}
\end{figure}

In Fig. \ref{fig:variance_v}, we show the standard deviation of the
drag force, $StdDev(F)$, as a function of the rotation rate, where
$StdDev(F)$=$\sqrt{\frac{1}{N}\sum_{i=1}^{N}(F_i-<F>)}$. $N$ is the
number of measurements in the force time series and $F_i$ is the $i$th
measurement. We note that the standard deviation is of the same order
of magnitude as its corresponding mean, and that it also increases
roughly logarithmically with the rate.

The slow increase of the mean drag force with rate appears to differ
from experimental observations in some previous studies, including
those by Wieghard and by Albert et al.
\cite{wieghard_75,albert_99}. In particular, Wieghard
\cite{wieghard_75} measured the drag force experienced by vertical
rods dipped into a rotating bed of fine dry sand. In this case, the
drag force had a weak dependence on the velocity: first decreasing
then increasing with increasing velocity. In the experiments by Albert
et al. \cite{albert_99}, the mean drag force on a cylindrical rod was
found to be independent of the drag velocity. 

In the case of Wiegard's experiments, the explanation for the
difference is relatively straightforward.  The velocity range used in
Wieghard's experiments is very different from both that used in our
and Albert et al.'s measurements. Wieghard investigated velocities
ranging from about 0.2 m/s to 2 m/s; the minimum of the drag force
appeared between 0.5 m/s and 1 m/s depending on the rod insertion
depth. Wieghard explained the variation of drag with speed in the
following way. The normal pressure and the frictional forces along the
slip surface provided resistance.  At lower speed, the inertial force
of the sand flowing around the body was small and negligible. When the
velocity increased, there was a reduction in drag because, presumably,
more contacts were slipping and kinetic friction is smaller than
static friction. At larger speeds, friction became less dependent on
the velocity; however, when the velocity was increased, an additional
inertial term led to an increase in the drag force. 

The velocities used in the Albert et al. experiments and in the
current experiments (of the order of 1 mm/s) are more comparable and
are much slower than that of Wieghard. To a first order approximation,
the present data is consistent with Albert's data, i.e. they both show
that the mean force is roughly independent of the velocity.  However,
we do see a slow, logarithmic increase in the mean force that differs
from the observation of Albert et al.

The explanation for this difference is not known, but it is
interesting to speculate on the cause.  Of course, there is the
obvious difference in dimensionality.  However, another difference
between the two experiments is that the present particles were softer (
a lower Young's modulus) than those used by Albert et al. In the
present experiments, the particles deformed elastically, whereas in
the experiments of Albert et al. an extrenal spring was deformed.  The
real issues include differences in the elastic time scales
vs. characteristic times for frictional events (e.g. creation and
destruction of force chains.)  and the amount of elastic deformation
of particles.  In this regard, we note the work by
Campbell\cite{campbell}.  Recent experiments by Hartley et
al. \cite{hartley_03} using the same type of particles as those of the
present experiments showed a qualitatively similar relation between
the mean force and the rate, albeit in a Couette system. These
experiments also showed that under static shear stresses, there was a
logarithmically slow relaxation of the force network.  Later in this
work, we will use a modified failure model inspired by this
observation to reproduce the slow increase in the mean drag force.

\begin{figure}[h]
\center{\parbox{6in}{\psfig{file=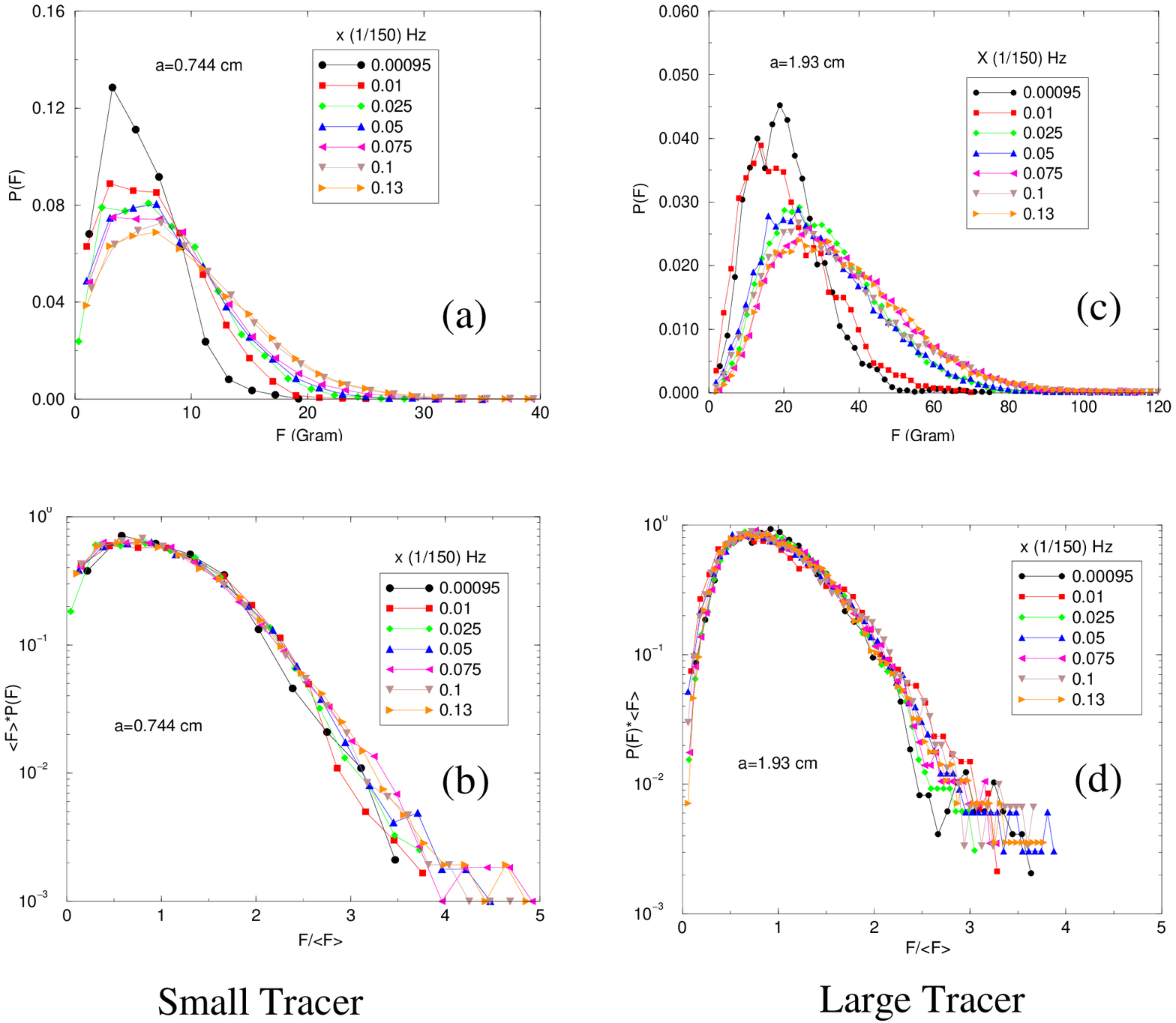,width=6in}}}
\caption{(a) Distributions of drag force at different rotation rates
for a tracer particle of size $a=0.744$ cm. (b) Same data as (a), but
with the horizontal axis rescaled by the corresponding mean force, and
the vertical axis multiplied by the mean force. The force
distributions collapse to a single curve. Note that since the vertical
axis in (b) is plotted on a logarithmic scale, the fall-off of the
distribution at larger forces is roughly exponential. (c) and (d) are
similar to (a) and (b), but for a larger tracer size $a=1.93$ cm.}
\label{fig:dist_f_v}
\end{figure}

In Fig. \ref{fig:dist_f_v}, we show drag force distributions for
different rotation rates. The left panel of Fig. \ref{fig:dist_f_v}
gives force distributions for a tracer particle of diameter $a=0.744$
cm, and the right panel gives data for $a=1.93$ cm. From
Fig. \ref{fig:dist_f_v}a, c, we note that, irrespective of the
particle size, the force distributions broaden and shift towards
larger forces as the rotation velocity increases. Interestingly, these
force distributions collapse into a single curve when scaled by the
corresponding mean force, as shown in Fig. \ref{fig:dist_f_v}b,
d. Thus, the mean force is one of the key control parameters for this
system. These data indicate a roughly exponential fall-off for large
forces, as seen in Fig. \ref{fig:dist_f_v}b, d, which shows the scaled
distributions on semi-log scales.  As the tracer size increases, one
noticeable change in the force distributions is that the probability
of very small force becomes smaller. An intuitive explanation is that
a larger tracer particle is more likely to be in contact with some
strong force chains at any time, thus reducing the probability of a
very small force.  This argument must be modified for tracers that are
much larger than the background particles.  As the tracer particle
diameter becomes very large, there are multiple contacts, some of
which involve strong force chains, and we expect that the distribution
for $F/<F>$ will no longer depend on the tracer diameter.

\subsubsection{Power Spectra and Correlations}

\begin{figure}[h]
\center{\parbox{4in}{\psfig{file=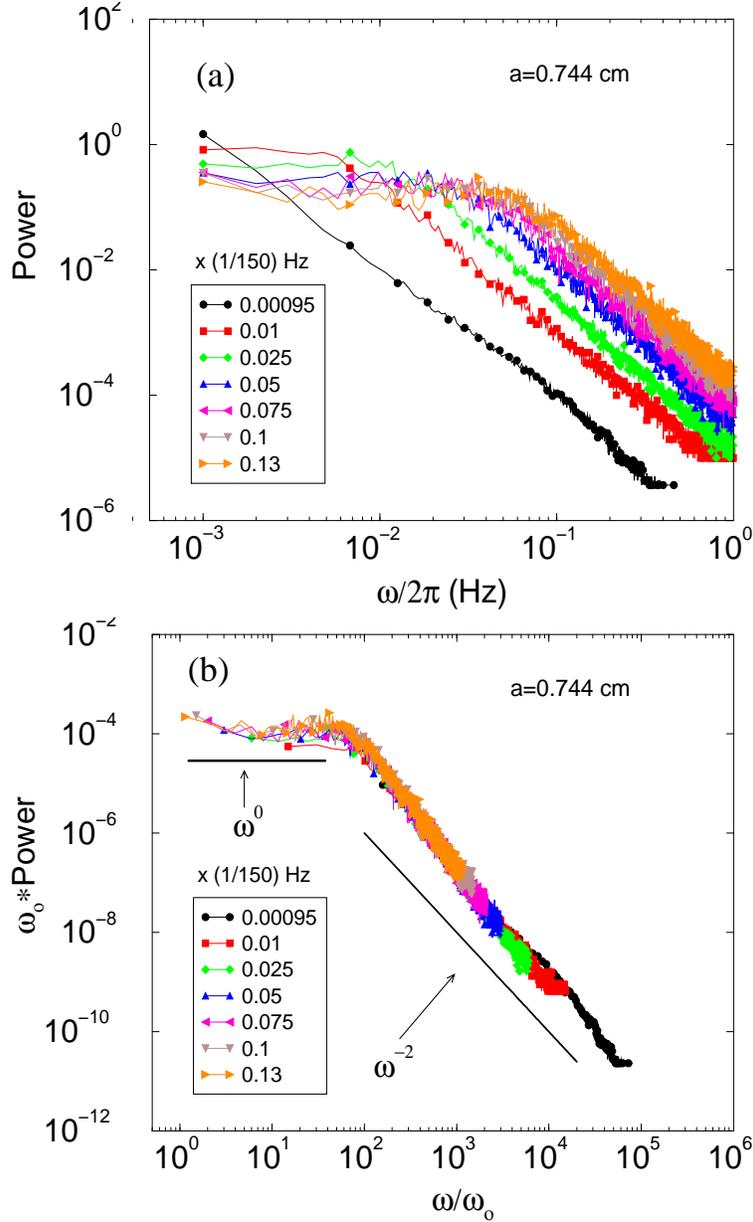,width=4in}}}
\caption{ (a) Power spectra, $P(\omega)$, from force time series at
different rotation rates. (b) The scaled power, $\omega_o P(\omega)$
is plotted against the scaled frequency, $\omega/\omega_o$, where
$\omega_o$ is the rotation rate. The data collapse nicely,
demonstrating rate independence. At large spectral frequency $\omega$,
the power spectra vary as $\omega^{-2}$, and at small frequency, the
power spectra are flat, suggesting that there is no correlation at
time scales larger than some constant factor of the inverse rotation
rate.}
\label{fig:power_spec_v}
\end{figure}

The power spectra, $P(\omega)$, resulting from such force time series
provide a useful quantitative measure of the relevant time scales for
force fluctuations. (Note that the mean force has been removed in
calculating the spectra.) In Fig.~\ref{fig:power_spec_v}a, we show
$P(\omega)$ vs. the frequency, $\omega$, on log-log scales. At high
frequency, the spectra fall off as $P(\omega) \propto
1/\omega^\alpha$, with $\alpha \simeq 2$. At low frequency, the
spectra vary more weakly, and are almost independent of the
frequency. The $1/\omega^2$ behavior at high frequency can be
explained by assuming a series of random jumps occurring on time
scales at least as fast as a crossover time $\sim 1/\omega^*$.  This
time corresponds roughly to the time for the tracer particle to travel
a few disk diameters. We will come back to this time scale below in
more detail. The power spectrum at low frequency is presumably
explained by the fact that there are no strong correlations at very
long time scales in the force time series.  A $1/\omega^2$ behavior
occurs in many other contexts, e.g. for frictional fluctuations
\cite{demirel_96} and stick-slip motions \cite{rozman_96}.

These spectra also show interesting rate invariance. In
Fig.~\ref{fig:power_spec_v}b, we rescale the power spectra data of
Fig.~\ref{fig:power_spec_v}a by dividing the $\omega$-axis by the
corresponding rotation rate, $\omega_0$, and multiplying $P$ by
$\omega_0$. This corresponds to rescaling time by $1/\omega_o$, or
alternatively by replacing time by angular displacement.
Fig.~\ref{fig:power_spec_v}b shows an excellent collapse of all the
data for the scaled power vs. the scaled frequency, and implies rate
invariance in the fluctuating component of the stresses.  Such
rate-invariance in stress fluctuations has also been observed by
Miller et al. \cite{miller_96} and Albert et al. \cite{albert_01}. An
argument for this rate invariance is provided in
Ref. \cite{behringer_01} which suggests that the system spends much of
its time in states close to static equilibrium, so that $\omega_0$
sets the time scale to move between states.

\begin{figure}[h]
\center{\parbox{4in}{\psfig{file=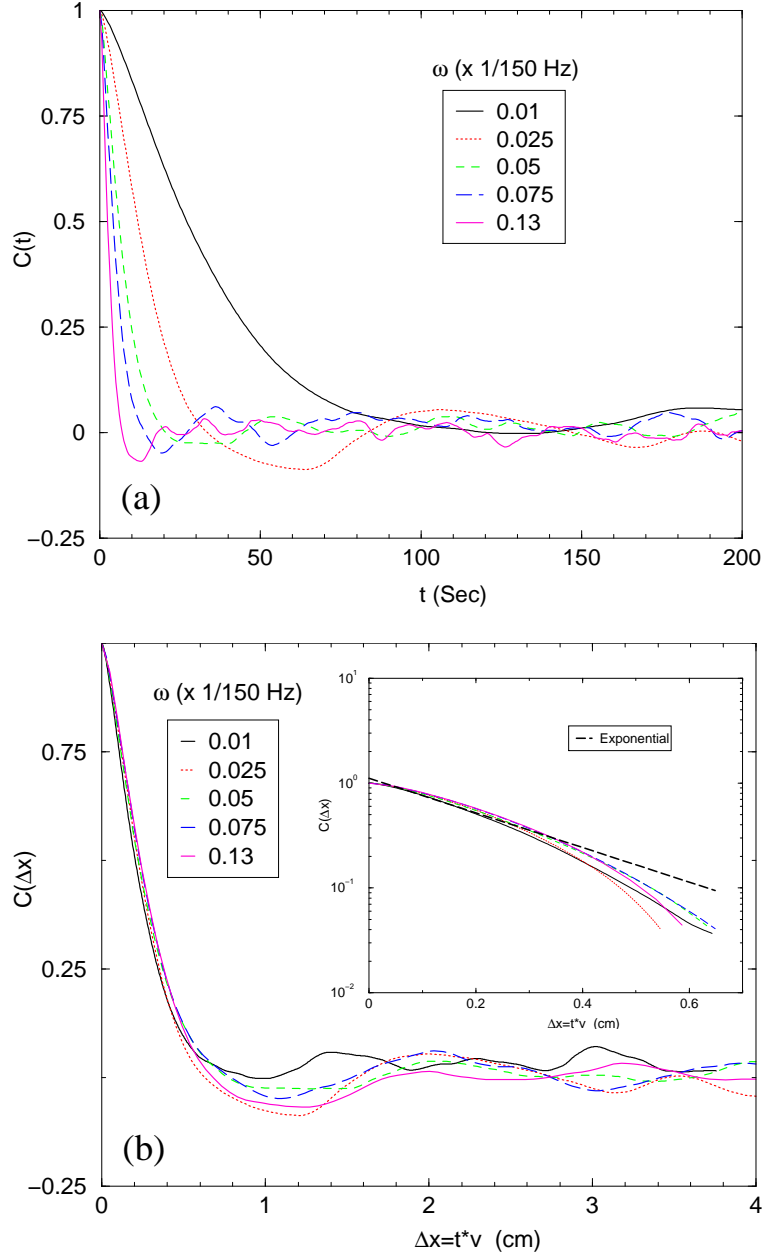,width=4in}}}
\caption{ (a) Correlation functions, $C(t)$, of the force time series
at different rotation rates. We note that correlation functions at all
rotation rates first drop quickly (exponentially) over a time scale of
$t_c$, then fluctuate around zero. (b) Rescaled correlation functions,
$C(\Delta x)$ v.s. $\Delta x$, where $\Delta x=v\Delta t = r \omega
\Delta t$. All rescaled correlation functions collapse to a single
curve, indicating a characteristic length scale, $\Delta x_c$. Note
that the correlation data and power spectrum data from
Fig.~\protect\ref{fig:power_spec_v} are related through
Wiener-Khinchin Theorem. Inset of (b) shows the correlation function
for small $t$'s, which corresponds to large frequency in the spectra.}
\label{fig:correlation}
\end{figure}

We can better understand the role of $\omega^{*}$ by calculating the
correlations resulting from these force time series. In
Fig. \ref{fig:correlation}a, we show correlation functions, $C(t)$,
for time series at different rotation rates (Note that $C(\Delta
t)=<F(t)F(t+\Delta t)>$, where the brackets denote an average of time,
and $C(\Delta t)$ is simplified as $C(t)$ when no confusion is
caused). These correlation functions generally drop quickly
(exponentially) to zero over a time scale of $t_c$, and then fluctuate
around zero, indicating that the signals are uncorrelated beyond that
time. If we rescale the data of Fig.~\ref{fig:correlation}a by
multiplying the $t$-axis by the corresponding velocities, all
correlation functions collapse to a single curve, as shown in
Fig. \ref{fig:correlation}b. The collapsed curve defines a
characteristic length scale, $\Delta x_c$, which is comparable to one
disk diameter. Intuitively, this can be explained by the fact that
force chains contacting the tracer particle tend to form and then fail
when the tracer particle moves by a few grain diameters, in agreement
with the characteristic length scale revealed in
Fig. \ref{fig:correlation}b.

We note here that the correlation data and the power spectra data are
a Fourier Transform pair according to the Wiener-Khinchin Theorem
\cite{nr_92}. Thus, the $1/\omega^2$ power spectrum at high frequency
can also be derived from the correlation data at small time
scales. Using the fact (inset of Fig.~\ref{fig:correlation}b) that the
correlation functions decay exponentially at early time as:
$C(t)=A_0\cdot exp(-t/t_c)$, the corresponding power spectrum can be
obtained by performing a Fourier transform:
\begin{eqnarray*}
P(\omega)&=&\int_{-\infty}^{\infty}C(t)exp(-i\omega t)dt\\
&=&\frac{2t_c}{1+(\omega t_c)^2}\\
&\approx&\omega^{-2}, \hspace{2mm} if\hspace{2mm} \omega \gg 1/t_c. \\
\end{eqnarray*}
\noindent Thus, for large frequency ($\omega \gg 1/t_c$), we expect
the power spectrum will decay as $1/\omega^2$.

\subsubsection{Avalanches and the Force Chain Force Constant.}

If we define an avalanche event to be a monotonic decrease in the
force time series, we can investigate the stress release process in
the system more quantitatively (similar results are found for the
stress build-up process). This approach is similar in spirit to the
approach of self-organized criticality (SOC) \cite{frette_96}, and it
is interesting to ask whether any sign of SOC is present in this
system.

\begin{figure}[h]
\center{\parbox{4.5in}{\psfig{file=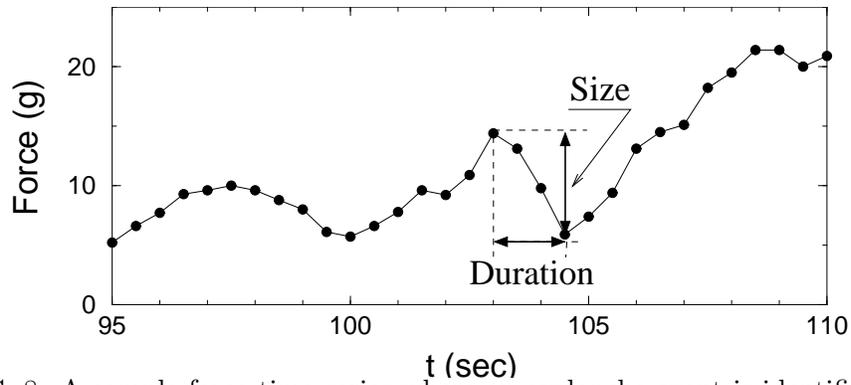,width=4.5in}}}
\caption[Illustration of an avalanche event]{A sample force time
series where an avalanche event is identified.}
\label{fig:ava_def}
\end{figure}

\begin{figure}[h]
\center{\parbox{4in}{\psfig{file=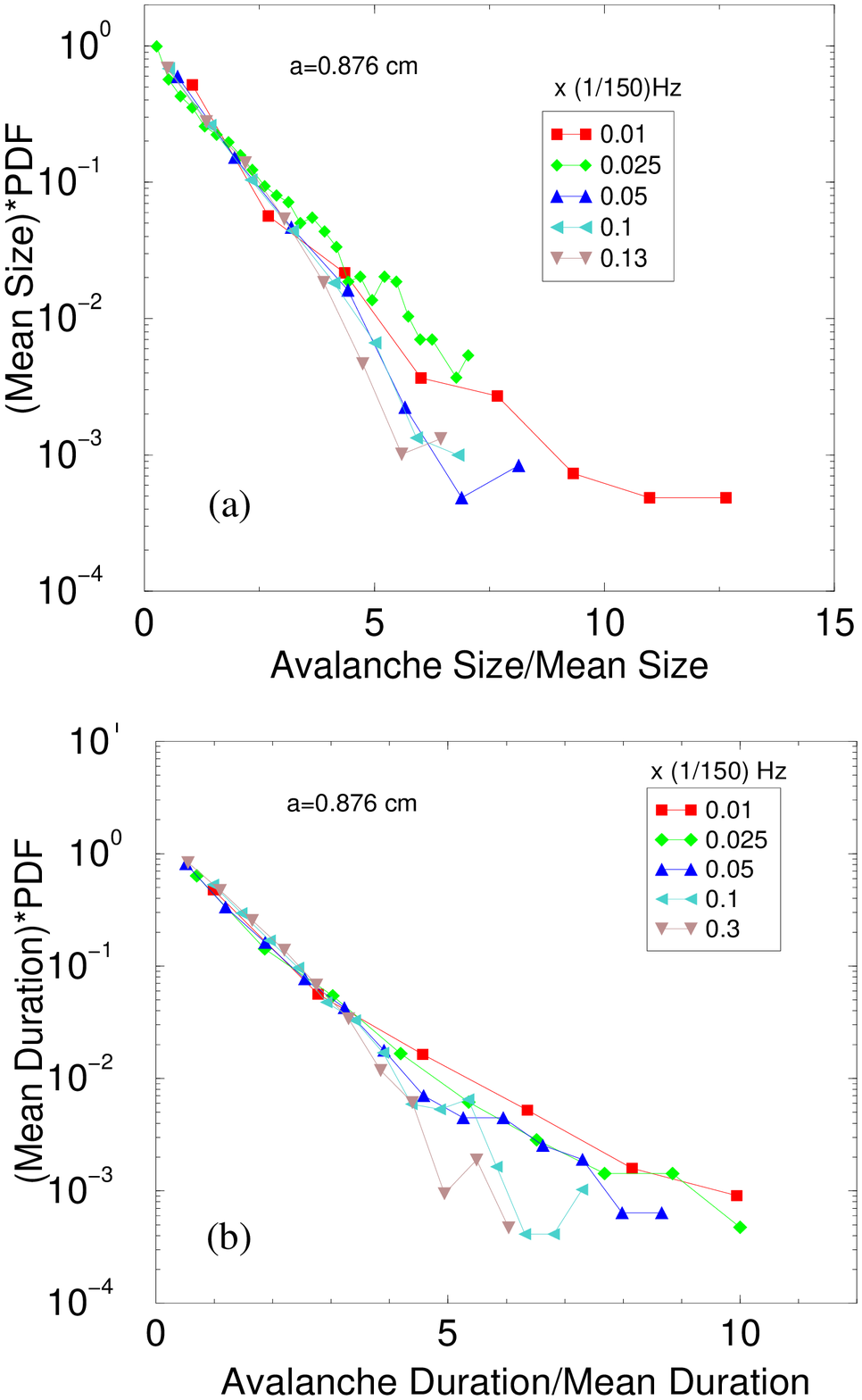,width=4in}}}
\caption{(a) Rescaled distributions of avalanche size, where the
horizontal axis is divided by the mean avalanche size and the vertical
axis is multiplied by the mean avalanche size.  (b) Rescaled
distributions of avalanche duration, where the horizontal axis is
divided by the mean avalanche duration and the vertical axis is
multiplied by the mean avalanche duration. These distributions
indicate an exponential decay of probabilities of finding large
avalanche sizes and durations.}
\label{fig:ava_size_dist}
\end{figure}

We denote the size of an avalanche to be the magnitude of the drop of
the force and the duration to be the time it takes for an avalanche
event to take place, as illustrated in Fig.~\ref{fig:ava_def}. With
such definitions, we can calculate the probability distributions for
both avalanche sizes and avalanche durations. We show such
distributions (properly rescaled) in Fig. \ref{fig:ava_size_dist} for
force time series obtained at different velocities. It is possible to
collapse all the distributions for avalanche size by dividing the
horizontal coordinate for each set of data by the corresponding mean
avalanche size and (and therefore necessarily multiplying the vertical
coordinate by the mean avalanche size). The avalanche duration
distributions are similarly rescaled by the corresponding mean
avalanche duration of each data set. In Fig. \ref{fig:ava_size_dist},
we show both data sets on log-lin scales, which emphasizes the roughly
exponential nature of the distributions. The flat tails at larger
values of the horizontal coordinates may be due to insufficient
statistics. These data suggest that there is a large probability of
finding small avalanche events in the system, while the probability of
finding a large avalanche event becomes exponentially small.  Note
that these distributions do not show any indication of power laws, as
one would expect for a self-similar process and SOC.

\begin{figure}[h]
\center{\parbox{4.0in}{\psfig{file=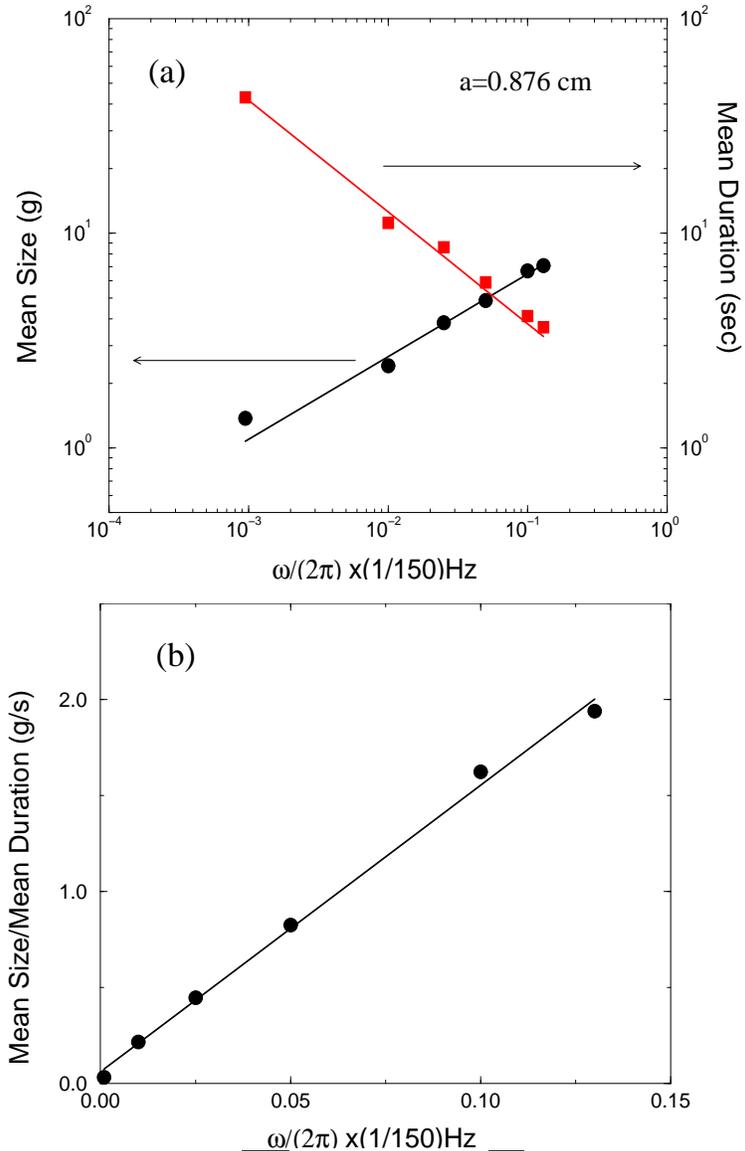,width=4.0in}}}
\caption[Mean avalanche size and duration vs. the rotation rate]{(a)
Mean avalanche size, $\overline{\Delta F}$, and duration,
$\overline{\Delta t}$, as functions of the rotation rate,
$\omega$. (b) The ratio of the mean avalanche size to duration,
$\overline{\Delta F}/\overline{\Delta t}$, as a function of $\omega$.}
\label{fig:mean_size_dura_v}
\end{figure}

It is interesting to ask how the mean avalanche size and duration
change with $\omega$. We show, in Fig.~\ref{fig:mean_size_dura_v}a,
data for the mean avalanche size, $\overline{\Delta F}$, and duration,
$\overline{\Delta t}$, as functions of the rotation rate. The mean
avalanche size increases with $\omega$ and the mean avalanche duration
decreases with $\omega$. Both the mean size and the and the mean
duration vary as power laws with $\omega$.  Particularly interesting
is the fact that the ratio of the mean avalanche size to duration,
Fig.~\ref{fig:mean_size_dura_v}b, also varies essentially linearly as
a power of $\omega$. The linear relationship between $\overline{\Delta
F}/\overline{\Delta t}$ and $\omega$ (or the medium velocity $v$)
suggests that there is an effective 'spring constant' for the force
chains, that can be defined as $\overline{\Delta F}/(v\overline{\Delta
t})$.  We develop this point further in the next few paragraphs.

\begin{figure}[h]
\center{\parbox{3.0in}{\psfig{file=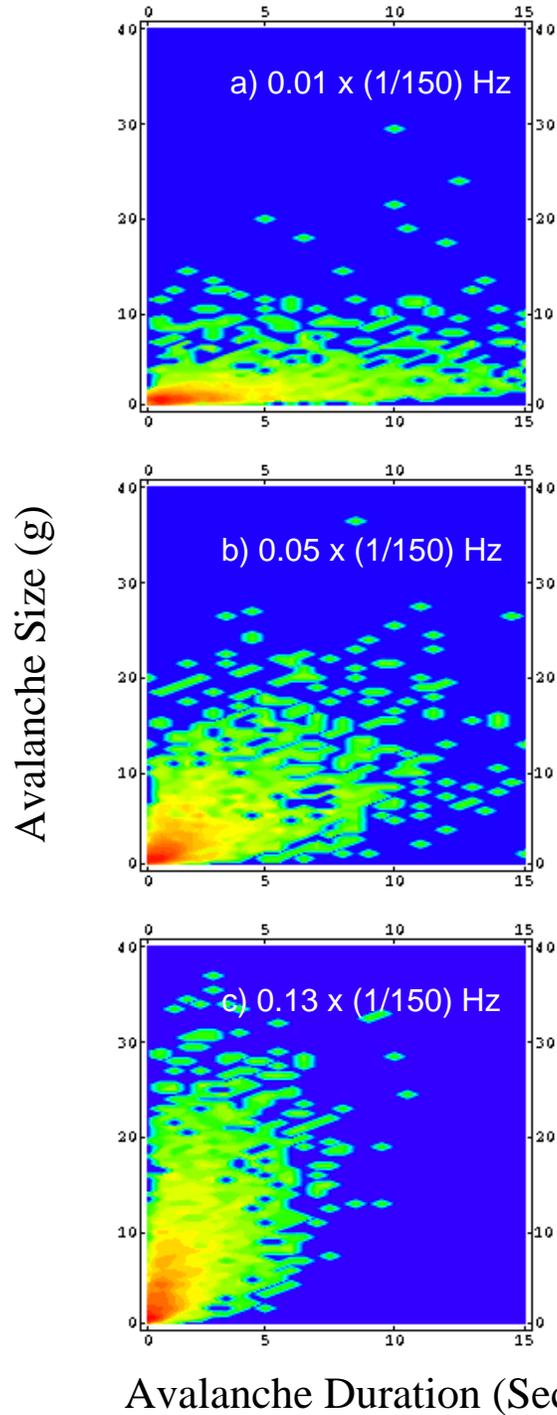,width=3.0in}}}
\caption{ 2D (greyscale representation) distributions of avalanche
size and duration for different rotation rates: (a) $\omega=0.01
\times (1/150)$ Hz, (b) $\omega=0.05 \times (1/150)$ Hz, and (c)
$\omega=0.13 \times (1/150)$ Hz. These distributions tend to be
largest along certain orientations/slopes. This slope increases when
the rotation rate is increased.}
\label{fig:ava_2D_dist}
\end{figure}

An obvious question is whether a large avalanche event (in terms of
its size) is in general associated with a longer duration, or perhaps
vice-versa. This question is addressed in Fig. \ref{fig:ava_2D_dist}
by calculating the 2D probability distributions for avalanches sizes
and durations. These distributions are given in
Fig. \ref{fig:ava_2D_dist}a-c for different drag velocities, using a
greyscale representation.  We see that these distributions are always
distributed around certain directions with positive slopes, which
suggests that, in general, a larger avalanche event lasts longer. We
also note that the slope of the distribution orientation increases
with increasing drag velocity. Based on the scalings of
Fig.~\ref{fig:ava_size_dist}, if we rescale the vertical and
horizontal axis in Fig. \ref{fig:ava_2D_dist} by the mean avalanche
size and mean avalanche duration, respectively, we expect that the
resulting distributions for different velocities would be peaked
around the same orientation. Indeed we have tested that this is the
case.

\begin{figure}[h]
\center{\parbox{4in}{\psfig{file=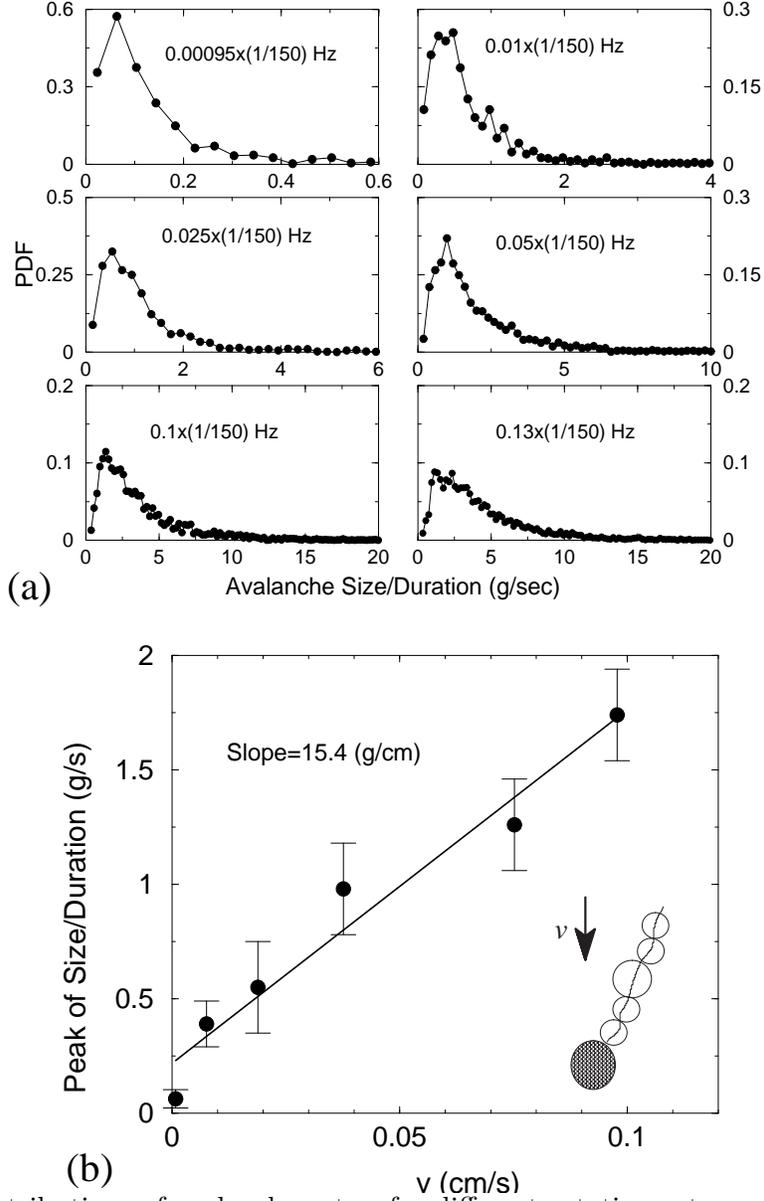,width=4in}}}
\caption{a) Distributions of avalanche rates, for different rotation
rates, where the avalanche rate is defined as the ratio of avalanche
size to duration, or roughly the 'slope' of an avalanche. (b) The
peaks in the avalanche rate distributions plotted against the rotation
speed. The slope of the resulting straight line fit gives an effective
force chain force constant, $k_{eff}=\frac{\Delta F}{\Delta
t}\frac{1}{v}=\frac{\delta F}{\Delta x}$. The inset shows a schematic
of = a force chain.}
\label{fig:force_chain_const}
\end{figure}

Since the 2D distributions for avalanche size and duration,
Fig. \ref{fig:ava_2D_dist}, tend to be oriented around a certain
direction, it is useful to consider an alternative approach to
characterize these events.  Namely, we define the avalanche rate to be
the ratio of the avalanche size and the corresponding duration, i.e.,
$Rate=\frac{Size}{Duration}=\frac{\Delta F}{\Delta t}$. We show the
distributions of rates for different medium velocities in
Fig. \ref{fig:force_chain_const}a. From this figure, we see first that
each distribution is peaked, which is consistent with our claim that
events have a most probable direction in Fig. \ref{fig:ava_2D_dist},
albeit with some spreading around that direction. Secondly, this
figure shows that when the rotation rate increases, the position of
the peak shifts to the right.

We extract the peak positions and plot them as a function of the
medium velocity, Fig. \ref{fig:force_chain_const}b. This figure shows
that the peak position increases roughly linearly with the medium
velocity. If we denote the slope of a least-squares linear fit to this
data as $k_{eff}$, then:
\begin{equation}
k_{eff}=\frac{\Delta F}{\Delta t}\frac{1}{v}=\frac{\Delta F}{\Delta x}.
\end{equation}
Thus, $k_{eff}$ resembles the force constant of a simple
spring. Indeed, Fig. \ref{fig:setup}c shows that the resisting forces
are mainly carried through chain-like structures, and one might
imagine that each of these force chains acts like a spring. The
collective force constant of these force chains is then rather well
defined, as suggested by the quantity, $k_{eff}$, extracted from
Fig. \ref{fig:force_chain_const}b. One has to keep in mind that since
Fig. \ref{fig:force_chain_const}b is obtained only for peak positions,
the actual effective force constant at a given instant can vary around
the $k_{eff}$ extracted here. A similar observation has been made in
Ref. \cite{kahng_01} by Kahng et al. concerning their 3D drag
experiment (see Fig. 2 in Ref. \cite{albert_00}). However, the force
constant revealed in those experiments reflects only the force
constant of the external spring.  That is, since it is much softer
than the effective spring constant of the grains, the force registered
on the force sensor is mainly due to the compression of the external
spring. By contrast, in our experiments, the effective force constant
gives a measurement of actual strength of the force chains in the
granular system.  Specifically, the force constant of the external
spring in our apparatus is much stronger than that associated with the
particles.

The above analysis supports the idea that force chains may be modeled
by springs as proposed in the model by Kahng et
al. \cite{kahng_01}. In Section IV below, we modify their model to
explain features of the data for the current experiments.

In the remainder of this section, we explore several other features of
the experimental results.

\subsection{Changing the Packing Fraction}

In this section, we describe experimental data and analysis associated
with changing the packing fractions in the system. For this set of
experiments, we fixed the rotation rate at $\omega_0=5.0\times
10^{-4}Hz$ and the tracer size at $a=1.25$ cm.

\subsubsection{Mean Drag Force and Force Distributions}

When we change the packing fraction, $\gamma$, we observe a
softening/strengthening transition similar to the one reported in Ref.
\cite{howell_99}.  Specifically, when $\gamma$ is below a critical
value, $\gamma_c$, the system is so loosely packed that it cannot
sustain force chains. In the regime $\gamma < \gamma_c$, when the
grains make contact with the tracer particle, they are almost
immediately pushed into open space, and no long-range force chains
form. On the contrary, when the packing fraction is above the critical
value $\gamma \geq \gamma_c$, there are always some force chains in
the bulk of the system, such as those shown in Fig.
\ref{fig:setup}c. In Fig. \ref{fig:force_series_gamma}, we show three
sets of force time series data obtained at different $\gamma$'s.  For
the data of $\gamma=0.561$, which is below $\gamma_c=0.645$, the
forces are close to zero, with a small amount of activity
corresponding to those events when the tracer particle makes contact
with grains. When $\gamma=0.653$, which is slightly above $\gamma_c$,
we already see more activity, and the average force signal increases
above the base line. When $\gamma$ is increased further, say to
$\gamma=0.754$, the force signal become much more active and the scale
of fluctuations is significantly larger.

Fig. \ref{fig:f_gamma}a shows the mean drag force as a function of the
global packing fraction $\gamma$. We identify two different regimes in
this figure. For smaller $\gamma$'s, the mean force can be fitted by a
linear function of $\gamma$: $F=a\gamma+b$, where a and b are
constants, while for larger $\gamma$'s, the mean force can be fitted
by a power-law, which parallels the results of Howell et
al. \cite{behringer_01}: $F=F_c+d(\gamma-\gamma_c)^\beta$, where d and
$\beta$ are constants.  We define $\gamma_c$ as the crossover value
from the linear to the non-linear regime.  In Fig. \ref{fig:f_gamma}b,
we show the mean force as a function of reduced packing fraction,
$r=\frac{\gamma-\gamma_c}{\gamma_c}$, for $\gamma \geq \gamma_c$ on
log-log scales to emphasize the power-law character in the nonlinear
regime.  In that regime, the exponent of the power law is
$\beta=1.53$.

In Fig. \ref{fig:dist_f_gamma}a, we show drag force distributions for
different packing fractions. As the packing fraction is increased, the
distributions widen and the means becomes larger, consistent with the
data of Fig. \ref{fig:f_gamma}. Again, if we rescale the force
distributions by the corresponding mean force, we obtain an
approximate collapse of all curves. Thus, the mean force is also the
appropriate scaling factor for the amplitude of the drag force
fluctuations.

Thus far, we have considered the mean properties and distributions of
the drag forces for different rotation rates and packing fractions. We
now combine these results and examine how the control parameters,
$\omega$ and $\gamma$, affect the drag force.

Fig. \ref{fig:dist_f_v_gamma} shows the combined drag force
distributions for various rotation rates and packing fractions. The
solid symbols are data for different $\omega$'s, and the open symbols
are data for different $\gamma$'s. All the distributions are rescaled
by their corresponding mean drag forces. Again, we see all rescaled
curves have nearly the same form.  This statistical invariance in the
force distributions is striking, since these data are obtained over a
wide range of rotation rates (more than two decades) and packing
fractions. This again confirms the key scaling role of the mean
force. We note too that these distributions decay roughly
exponentially for large forces, in the spirit of the q-model
\cite{q-model}.

For a given tracer particle, changing the rotation rate or changing
$\gamma$ both affect the mean drag force, although the former is only
a weak effect. In Fig. \ref{fig:f_v_gamma}a, we combine the data for
mean drag forces from Fig. \ref{fig:f_v} and \ref{fig:f_gamma} in a
single plot, where the top axis is the rotation rate, $\omega$, and
the bottom axis is the reduced packing fraction
$r=(\gamma-\gamma_c)/\gamma_c$. When $\gamma$ is fixed, the mean force
(solid circles) increases slowly with $\omega$, where this slow
increase is adequately described as a logarithm. When $\omega$ is
fixed, the mean force (solid squares) increases rapidly with $\gamma$,
and this increase is described by a power-law.  If we assume that
$\bar{F}$ can be written in a product form as
$\bar{F}=f_1(a)f_2(\omega)f_3(r)$, for our given tracer particle size,
we find that a good description of the data is given by:

\begin{equation}
\bar{F}=\frac{1}{14.51}(22.802+2.588\log\omega)(2.502+174.91 r^{1.529}).
\end{equation}

\noindent
Fig. \ref{fig:f_v_gamma}b shows the mean drag force $\bar{F}$ in a 3D
perspective plot. From this figure, we see that an increase of the
rotation rate, $\omega$, leads to an increase of the mean drag force,
qualitatively resembling what occurs due to an increase in the packing
fraction, $\gamma$, but on a much weaker scale. Similar effects on the
stress due to changes in the shear rate and packing fraction were also
observed in a 2D granular Couette systems \cite{hartley_03}.

We also examine how the diameter of the tracer particle, $a$, affects
the mean drag force. In Fig. \ref{fig:f_a}, we show the mean drag
force as a function of the tracer diameter for different rotation
rates at a given packing fraction $\gamma=0.754$. From these data, we
see that the increase in the mean force with tracer particle size is
faster than linear.

It is interesting to contrast these results with what one would expect
for a particle, typically much larger than a molecule, that is moving
through a viscous fluid.  According to Stokes's law \cite{pathria_96},
the drag force, is proportional to the diameter of the tracer
particle, the coefficient of viscosity of the fluid, and the relative
velocity of the fluid and the tracer.

It is also interesting to compare our results to the experiments by
Albert et al.\cite{albert_01} on drag through a granular material.  As
noted, these authors observed rate independent forces.  They also
found a linear dependence of the drag force on the diameter of the
drag rod.  However, it is perhaps not surprising that in the present
experiments the diameter dependence of the drag force is nonlinear,
since the tracer particle size is comparable to the size of
surrounding grains (the maximum size ratio is 2.6), unlike the
situation in the experiments of Albert et al.

\subsubsection{Rescaling of Power Spectra and Avalanches}

In Fig. \ref{fig:power_spec_gamma}a, we show power spectra of force
time series for different packing fractions. In this case, variations
of the power spectra with $\gamma$ are qualitatively similar to those
due to changes in the rotation rate shown in
Fig. ~\ref{fig:power_spec_v}a, although the magnitude of the changes
with $\gamma$ is much greater.  It is interesting to rescale these
spectra to see if they will collapse onto a common curve. In this
regard, we note from Parseval's Theorem \cite{nr_92}, that the
integral of the power spectral density over frequency is equal to the
mean square amplitude of the signal, i.e.,

\begin{equation}
\frac{1}{2\pi}\int_{-\infty}^{\infty}P(\omega)d\omega=\frac{1}{2\pi}\int_{-\infty}^{\infty}|F(\omega)|^2d\omega=
\int_{-\infty}^{\infty}|f(t)|^2dt,
\end{equation}

\noindent
where $F(\omega)$ and $f(t)$ are a Fourier pair.  Hence, the integral
of the power spectrum, which is proportional to the mean square
amplitude of force signals, $<f^2>=\int_{-\infty}^{\infty}|f(t)|^2dt$,
can be used as an appropriate scale factor for the spectra in
Fig. \ref{fig:power_spec_gamma}a. Indeed, when these spectra are
normalized by the corresponding $<f^2>$, we obtained a good collapse
of data, as shown in Fig. \ref{fig:power_spec_gamma}b. Additionally,
we show the scaling factor, $<f^2>$, vs. the reduced packing fraction,
$r$, in Fig. \ref{fig:f2_reduced_gamma}. These data can also be fitted
to a power law, and the exponent is almost twice as large as the
exponent associated with the power-law for the mean force,
Fig. \ref{fig:f_gamma}.

Before turning to the model, we note that the avalanche data
calculated from force time series for different packing fractions are
similar to those for different rotation rates. We have tested that
distributions for both avalanche size and duration decay
exponentially, and can be rescaled by the respective mean avalanche
size and duration to obtain good collapse of the data.

\section{Model and simulations}

In this section, we turn to a stochastic failure model, based on one
originally proposed by Kahng et al. \cite{kahng_01} to understand the
experimental data of Albert et al.\cite{albert_99,albert_00}.  We
modify this model appropriately to account for several features that
are unique to the present 2D granular system.  Specifically, we make
two modifications to the original model:\\ \indent {\bf i)} First, we
allow the band of thresholds to be wide enough so as to generate
random force patterns, and we use exponentially distributed thresholds
to produce more realistic force distributions; \\ \indent {\bf ii)}
Second, we introduce a time-dependent threshold to explain the slow
(logarithmic) increase of the mean drag force with the rate.\\ \indent
We also note that since the particles are only one layer deep in the
2D experiments, we do not need any depth dependence.  In the reminder
of this section, we first briefly introduce the basic model.  We then
make modifications to the model and perform simulations to compare
with the present experimental data.

\subsection{The Original Spring Model}

The original model was constructed to simulate the drag force
experienced by a vertical cylinder inserted to a given depth in a
granular bed \cite{kahng_01}.  In this model, the grains move with
constant speed $v$ in the x-direction, and the tracer particle is
simply represented by a block, as shown in
Fig.~\ref{fig:model_cartoon}a.  The tracer particle interacts with
grains that are assumed to be supported by force chains.  The
particle-tracer interactions are modeled as linear springs with a
force constant $k_0$, where there are $n$ such springs.  (The
assumption of a single spring constant is in part justified for the
present data by the analysis of an effective force chain force
constant, $k_{eff}$, in the experimental data as in
Fig.~\ref{fig:force_chain_const}.)  Necessarily, the spring constant,
$k_{eff}$, refers to the collective mean response, instead of a force
constant for an individual force chain.  As time advances, each spring
is compressed by an amount $\Delta x$, which is determined by the
velocity $v$ and by $\Delta t$, the time interval over which
compression has occurred, i.e.

\begin{equation}
f=f_0+k_0\Delta x =f_0+k_0v\Delta t,
\end{equation}

\noindent where, $f_0$ is a small initial force proportional to the
local pressure in the system. This is illustrated in
Fig.~\ref{fig:model_cartoon}b. At $t=0$, a spring makes contact with
the tracer particle, corresponding to the formation of a force
chain. The spring is then compressed as time advances. If the spring
(force chain) is too compressed, e.g., the force $f$ exceeds a
threshold, $g$, the spring fails, and the force on the spring is
relaxed to $f_0$.  In addition, the threshold $g$ is updated to a new
value chose at random from an appropriate distribution.  In the
original model, $g$ was uniformly distributed over an interval [$g_0$,
$g_1$]. Over time, the process of spring compression (force chain
formation) and failure continues. At any given time, the drag force is
the sum of the forces from all $n$ springs.

The original model \cite{kahng_01} also assumes that the effective
force chain springs are much stronger than the external spring
associated with the machine that is pushing the tracer.  In such a
case, the drag force, which is typified by stick-slip dynamics, is a
function of the strength of the external spring. Khang et al. focused
on the stick-slip regime, since this corresponded to what was observed
in the 3D drag experiments by Albert et al.

However, in the present experiments, the effective spring constant of the
drive is significantly larger than that of the particles.
Consequently, we do not observe stick-slip behavior, but rather
random force fluctuations.  We must take into account this
different feature of our experiments, and we
now turn to appropriate modifications of the model.

\subsection{Modification I: Wide Threshold Bands and Exponentially Distributed Thresholds}

\begin{figure}[h]
\center{\parbox{5in}{\psfig{file=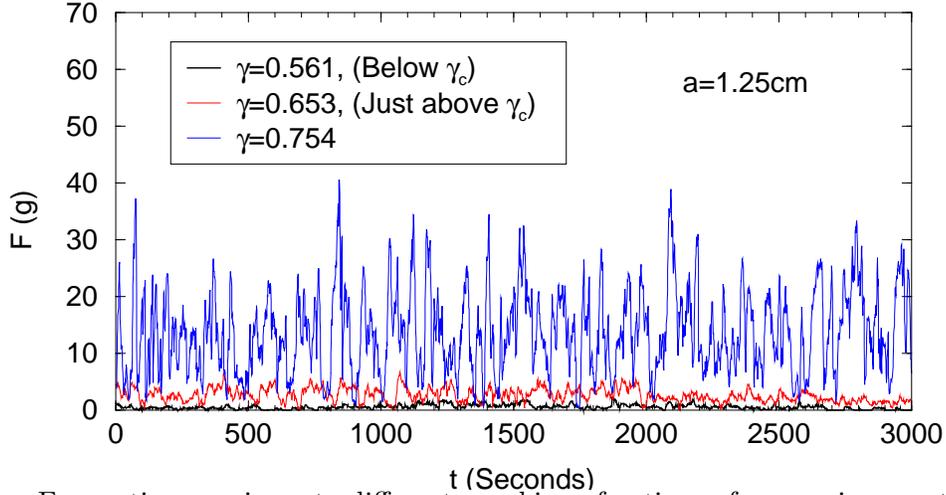,width=5in}}}
\caption{Force time series at different packing fractions for a given
rotation rate ($\omega=0.075 \times (1/150)$ Hz) and a given tracer
size ($a=1.25$ cm).  There exists a critical packing fraction,
$\gamma_c$, similar to that found in Ref. \protect \cite{howell_99},
(see text). Below $\gamma_c$, the force is relatively small and the
friction between particles and the bottom plate is comparable with the
contact force between particles; above $\gamma_c$, enduring contact
forces dominate and force chains form in the bulk of the system,
leading to strong fluctuations in the force time series. The mean drag
force increases rapidly when the packing fraction is increased.}
\label{fig:force_series_gamma}
\end{figure}

\begin{figure}[h]
\center{\parbox{4in}{\psfig{file=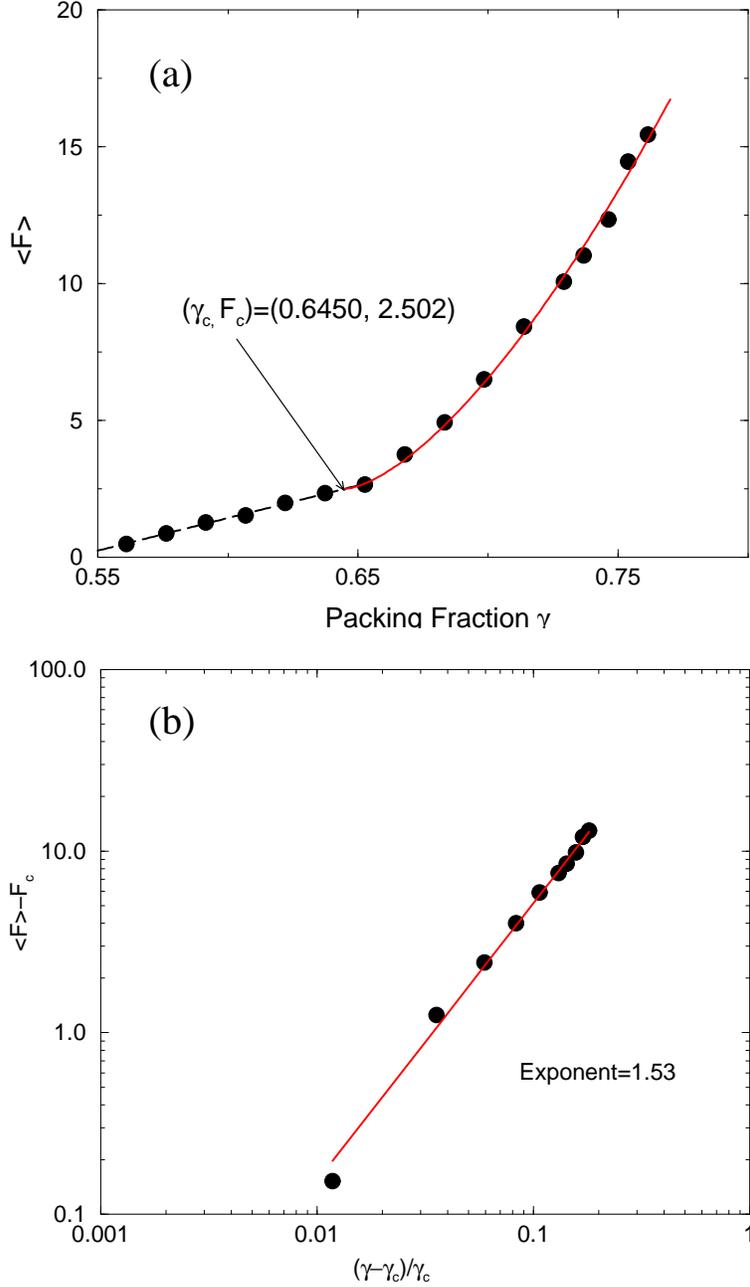,width=4in}}}
\caption{(a) The mean force, $<F>$, as a function of the packing
fraction, $\gamma$, for a tracer particle of size $a=1.25$ cm. When
$\gamma$ is below $\gamma_c$, the mean force increases linearly with
$\gamma$; when $\gamma$ is above $\gamma_c$, the force increases like
a power-law, as shown in (b) The mean force as a function of reduced
packing fraction, $r=\frac{\gamma-\gamma_c}{\gamma_c}$ for packing
fractions greater than $\gamma_c$, on log-log scales.}
\label{fig:f_gamma}
\end{figure}

\begin{figure}[h]
\center{\parbox{4in}{\psfig{file=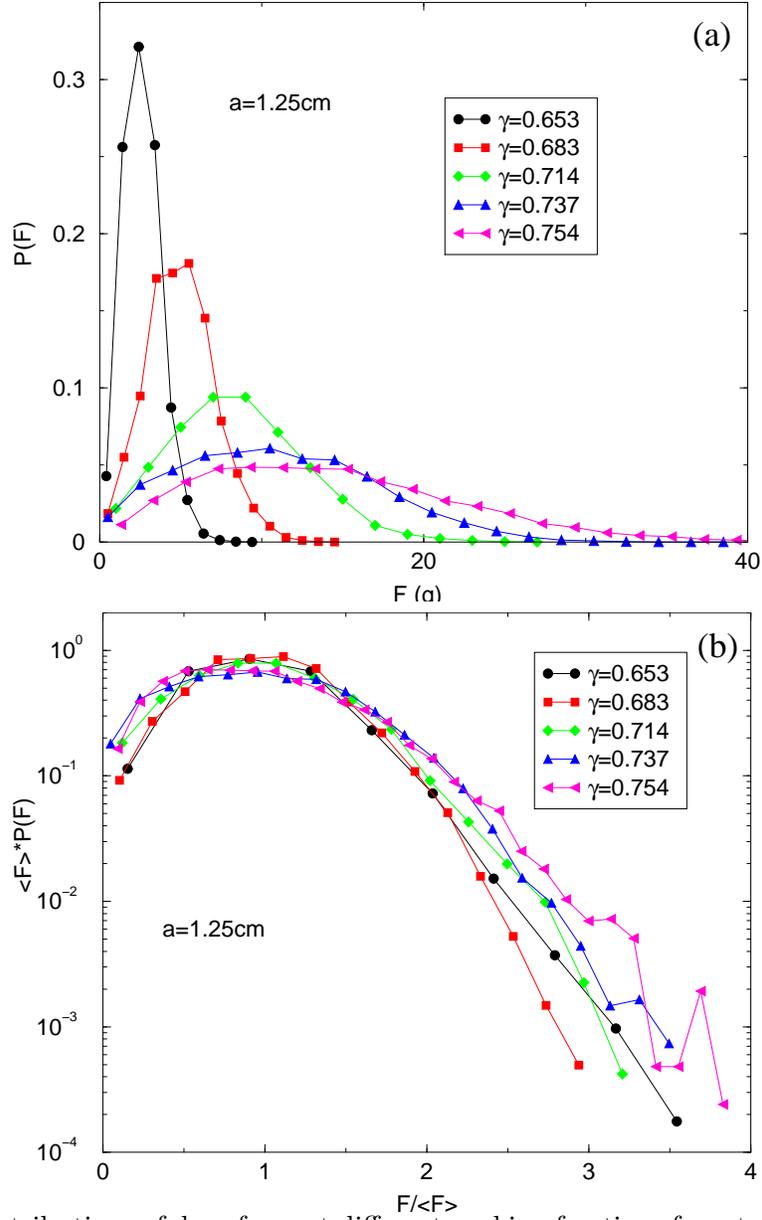,width=4in}}}
\caption{(a) Distributions of drag force at different packing
fractions for a tracer particle of size $a=1.25$ cm. (b) Same data as
(a), but with the horizontal axis rescaled by the corresponding mean
force, and the vertical axis multiplied by the mean force. Force
distributions collapse reasonably well onto a single curve.}
\label{fig:dist_f_gamma}
\end{figure}

\begin{figure}[h]
\center{\parbox{4in}{\psfig{file=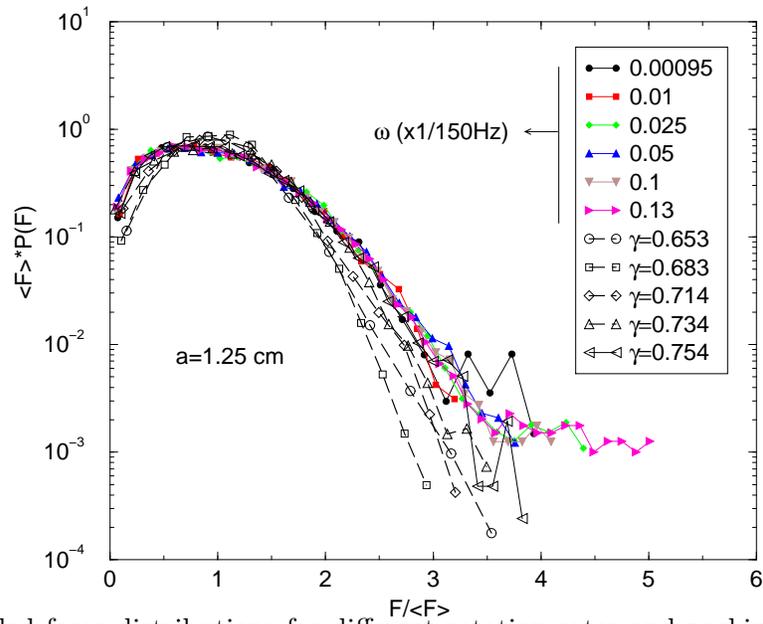,width=4in}}}
\caption{Rescaled force distributions for different rotation rates and
packing fractions.  These distributions collapse reasonably well onto a
single curve, suggesting a strong statistical invariance after
rescaling by the mean force.}
\label{fig:dist_f_v_gamma}
\end{figure}

\begin{figure}[h]
\center{\parbox{4in}{\psfig{file=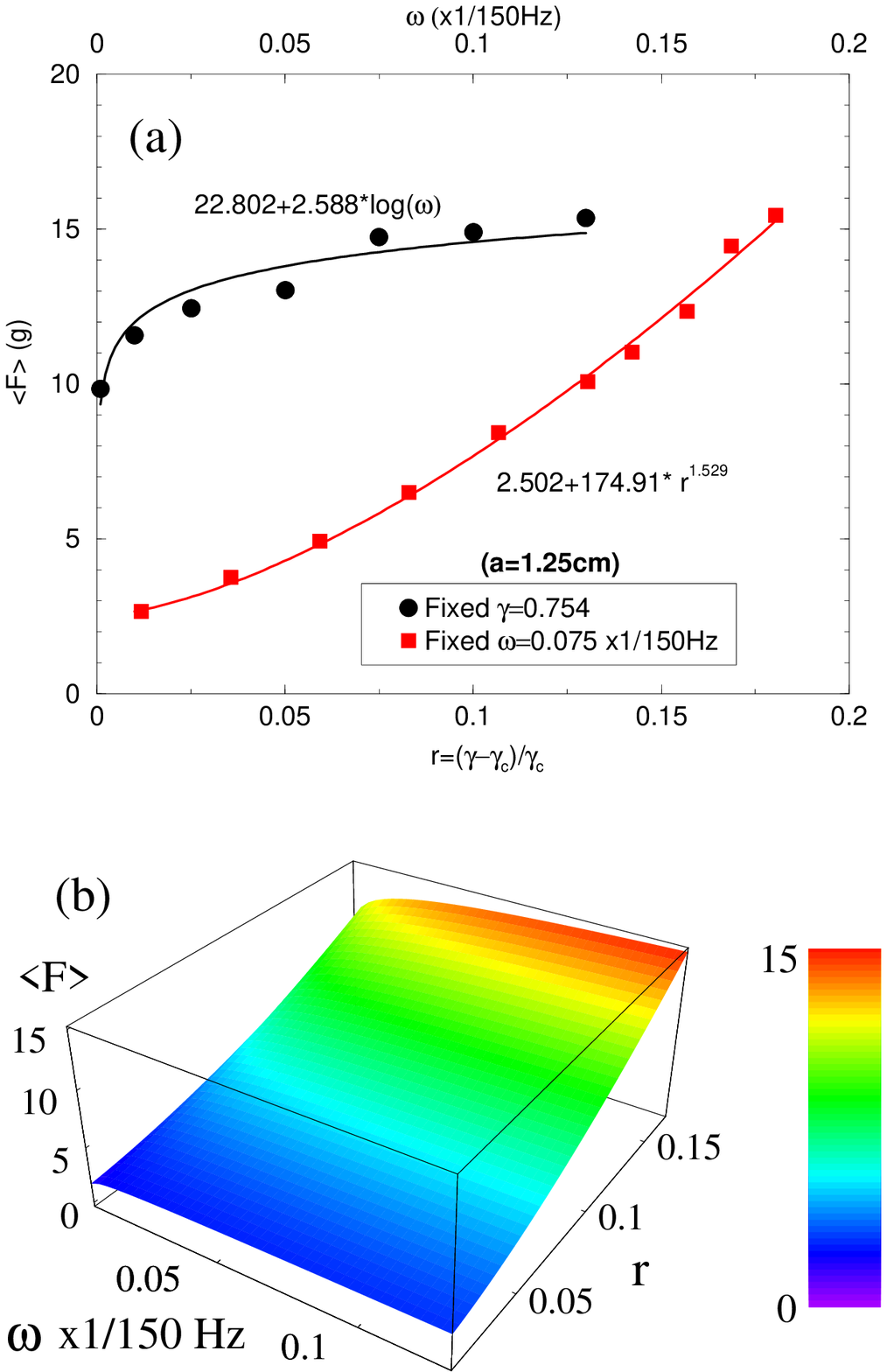,width=4in}}}
\caption{(a) The mean force as a function of drag velocity and medium
packing fraction for a given tracer particle. The solid symbols are
experimental data; the lines are a logarithmic fit and a power law fit
respectively. (b) With fits obtained from (a), we plot in b) a 3D
perspective plot showing how the mean force changes in the parameter
space formed by $\omega$ and $\gamma$. An increase of $\omega$ tends
to have the same effect on the mean force as an increase of $\gamma$,
although the former effect is much weaker.}
\label{fig:f_v_gamma}
\end{figure}

\begin{figure}[h]
\center{\parbox{4in}{\psfig{file=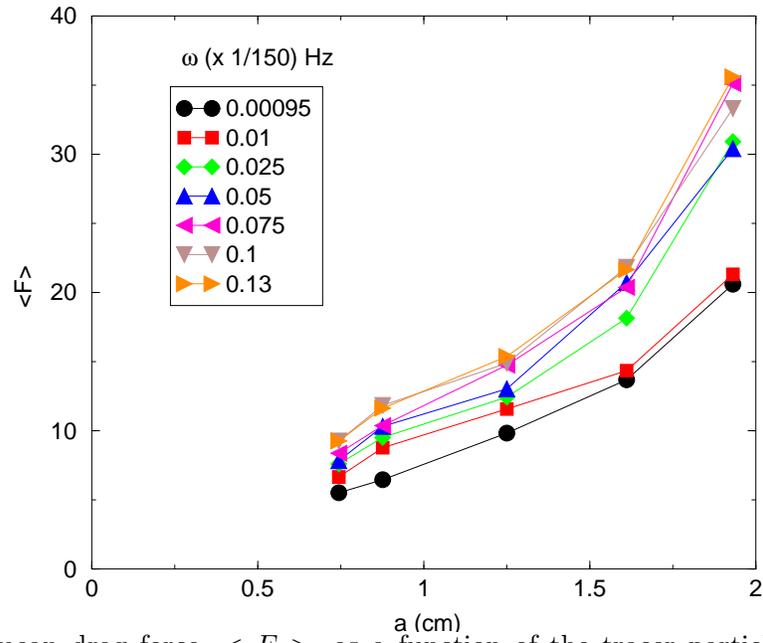,width=4in}}}
\caption{ The mean drag force, $<F>$, as a function of the tracer
particle diameter, $a$, for different rotation rates.}
\label{fig:f_a}
\end{figure}

\begin{figure}[h]
\center{\parbox{4in}{\psfig{file=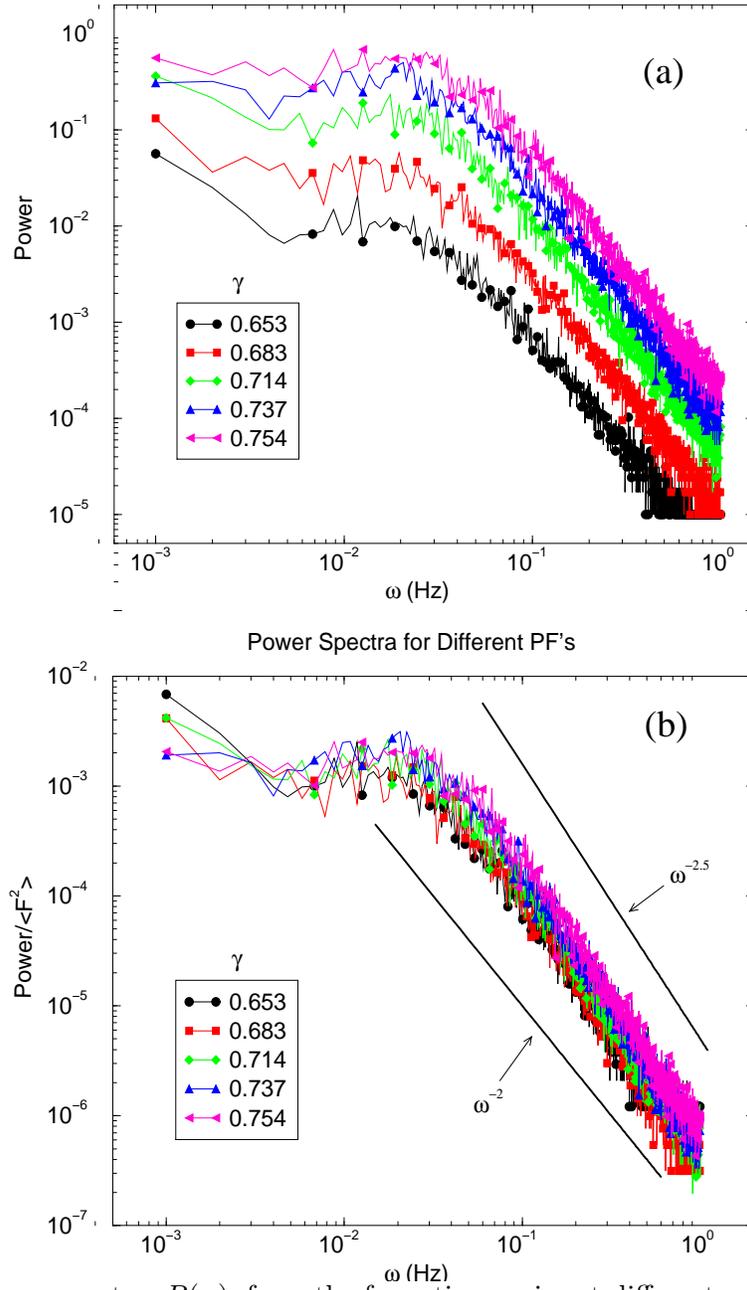,width=4in}}}
\caption{ (a) Power spectra, $P(\omega)$, from the force time series
at different packing fractions. (b) Scaled power spectra, with the
vertical axis divided by the mean square amplitude of the signal.}
\label{fig:power_spec_gamma}
\end{figure}

\begin{figure}[h]
\center{\parbox{4in}{\psfig{file=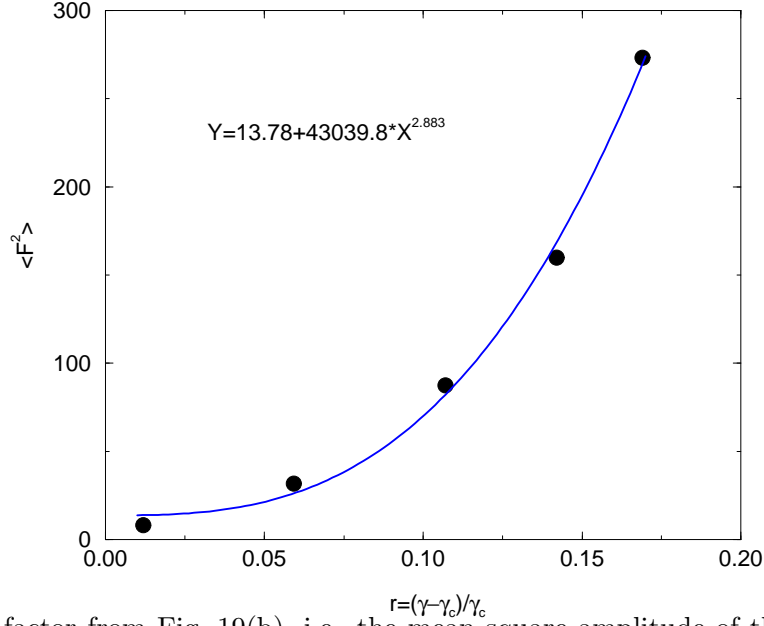,width=4in}}}
\caption{ Scale factor from Fig.~\protect
\ref{fig:power_spec_gamma}(b), i.e. the mean square amplitude of the
signal, $<F^{2}>$, vs. reduced packing fraction r. These data again
can be fitted by a power law.}
\label{fig:f2_reduced_gamma}
\end{figure}

\begin{figure}[h]
\center{\parbox{4in}{\psfig{file=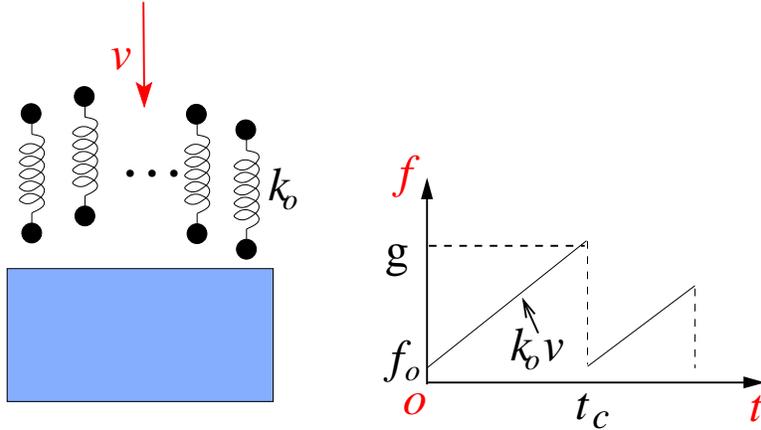,width=4in}}}
\caption{ (a) Schematic drawing of the failure model: the block is a
simplified representation of the tracer particle.  Force chains in the
bulk of the granular system opposing the tracer particle are modeled
by springs with a force constant $k_0$. The tracer particle is always
in balance with the collection of opposing springs/force
chains. (b). The force from each spring increases linearly with time,
until it fails at the point where the force reaches a threshold value,
$g$.  In the physical experiment, this occurs due to the slippage
between the tracer particle and the grains, or among grains
themselves. After the failure of a spring, the force on the spring is
reset, and the threshold is updated with another value drawn from the
distribution of thresholds and the process continues.}
\label{fig:model_cartoon}
\end{figure}

We begin by considering the effect of the width of the threshold band
[$g_0$, $g_1$]. As one would expect, this width qualitatively affects
the drag force patterns. When the threshold band is narrow, as in
Fig.~\ref{fig:force_series_gamma}a, for $[g_0, g_1]=[0.49, 0.51]$, the
force time series exhibits a regular sawtooth pattern. This is because
all the springs fail almost at the same time, resulting in a regular
pattern of buildup and release. When the threshold band is wider, the
force pattern becomes more random (e.g., $[g_0, g_1]=[0.1, 0.9]$ ).
This more closely resembles what occurs in the the present
experiments.

However, if the threshold $g$ is uniformly distributed between [$g_0$,
$g_1$], the resulting force distributions are symmetric with respect
to the mean drag force, as shown in Fig.~\ref{fig:force_series_gamma}b
for a 10 spring system. The symmetry of this distribution differs from
those of the experiment, and simply reflects the symmetry of the
failure distribution. The data of avalanche size distribution in
Fig. \ref{fig:ava_size_dist} suggest that the probability of finding a
large event becomes exponentially small. Thus, it is reasonable to
assume that the distribution of $g$'s is likewise exponential. We
expect that most of the time, the force chains break at small forces,
and only in rare events, do the force chains survive to reach a large
threshold. Using this assumption, we obtain a force time series such
as that shown in Fig.~\ref{fig:force_series_gamma}c.  We show the
resulting force distributions (for 10 springs) in (d).  In contrast to
Fig. \ref{fig:force_series_gamma}b, these new force distributions
obtained with exponentially distributed thresholds are significantly
closer in appearance to the experimental data, as in
Fig. \ref{fig:dist_f_gamma}.  

Note that the mean force in the model is found by summing over $n$
independent variables, $x_i$, where $x_i$ is the compression of spring
$i$.  The mean value of any one of these is then $\bar {x_i} =
(1/2)\bar {g}$, where $\bar{g}$ is the mean determined from the
distribution of $g$'s.  As $n$ grows, we expect that the distribution
of total force $F$ will approach a Gaussian with a mean value $n \bar
{g}$ and a width $\sqrt{n} \sigma$, where $\sigma^2$ is the variance
of $g$.  Indeed, the statistical properties of the model follow from
the fact that the force is a sum over $n$ uncorrelated random
variables where the maximum of each variable is drawn from the
appropriate distribution of $g$'s.

Apart from the force distributions, for other aspects of the simulated
data (power spectra, distributions of avalanche size/duration, and
force chain force constants), uniformly distributed thresholds do not
lead to significantly different results than thresholds that are
exponentially distributed, as long as the threshold band is wide
enough. Below, we will focus on the simulated data derived from
exponentially distributed thresholds.

In Fig.~\ref{fig:model_power_spec}, we show power spectra and their
rescaled form for different velocities calculated from the
model. These data are in remarkable qualitative agreement with the
experimental data shown in Fig.~\ref{fig:power_spec_v}.

In Fig.~\ref{fig:model_ava_size_dist}, we show in (a) the
distributions of avalanche sizes derived from the model simulations
and in (b) the rescaled distributions of avalanche durations derived
from the model simulations. Both distributions of avalanche size and
duration are roughly exponential for large arguments, as are the
experimental data, Fig. \ref{fig:ava_size_dist}.  Note, however, that
the size distributions in this figure are not rescaled while those in
Fig.~\ref{fig:ava_size_dist}a are.

Similarly, in Fig.~\ref{fig:model_force_chain_const}, we show
avalanche rate distributions at different velocities in (a) and the
derived effective force chain force constant in (b). This figure
compares well with the experimental data shown in
Fig.~\ref{fig:force_chain_const}. The effective force chain force
constant from the simulation data is $k_{eff}=30.7$, which is of the
same order of magnitude as $nk_0$, where $n=10$ (the number of
springs) and $k_0=1$ (the individual force constant of a spring).

\subsection{Modification II: Decaying Thresholds}

The model so far has been able to reproduce a number of experimental
observations. However, if we calculate the mean drag force, $<F>$, as
a function of the medium velocity, $v$, we find that $<F>$ is
independent of $v$, as shown in
Fig.~\ref{fig:model_f_v_no_decay}. Fig.~\ref{fig:model_f_v_no_decay}a
shows force distributions for several different velocities, and they
all fall on the same curve, with almost the same mean and
variance. Fig.~\ref{fig:model_f_v_no_decay}b is a direct plot of mean
drag force as a function of velocity, which shows a rate-independent
result. This differs from the experimental finding that the
mean drag force increases logarithmically with the velocity.

The fact that the model is rate-independent is not surprising.  The
instantaneous force state is found by summing over the $n$ springs.
The state of each spring does not depend on the velocity of the block,
but only on the displacement of the block since it was last reset to
$f_o$.  In such a displacement-controlled system, there can be no
velocity dependence.  

One possible way to account for the rate-dependence is to recognize
that there is failure of some contacts due to creep, and we explore
that possibility here.  In this regard, we note recent work by Hartley
et al. (Fig. 2 in \cite{hartley_03}) involving similar particles to
those used here.  These authors reported logarithmically slow
relaxation of the force chain network in their 2D granular Couette
system. Specifically, in these experiments, 2D photoelastic grains
were sheared steadily so as to establish a strong force chain network.
The shearing was abruptly stopped and the particle-scale forces in a
section of the Couette annulus were monitored thereafter. The force
chains relaxed (became weaker) over many hours, with the total stress
in the system decaying logarithmically slowly, presumably due to the
collective rearrangements of the grains and failure under creep at
contacts that were near to failure. Such failures became progressively
more difficult over time because, presumably, the contacts near
failure became less numerous, and also perhaps due to geometric
constraints on successive rearrangements.

To make a connection with the model, we note that one interpretation
of the Hartley et al. experiments was that the force chains become
logarithmically weaker over time, which means that the threshold of
each spring should decrease with time. This is illustrated in
Fig. \ref{fig:decay_g}. For two processes with different velocities
($v_1 > v_2$), if the originally chosen thresholds for a spring are
$g$ in each case, by the time a spring reaches its failure point, this
threshold has become smaller. Since $v_2 < v_1$, by the time failure
actually occurs, the threshold for the slow process ($v_2$) is smaller
than that of the fast process ($v_1$).  The longer one waits, the
smaller the threshold. Hence, we assume the threshold, $g$, is
time-dependent and decreases logarithmically with a time constant
$t_0$:

\begin{equation}
%g(t)=1-A \log( t/ t_0).
g(t)=1-\frac{\log t}{\log t_0},
\end{equation}

\noindent where, $t_0$ is a large value (about $10^5$ times the time
step) that sets the slow relaxation time scale/amplitude.

With such a decaying threshold, $g(t)$, we recalculate the drag force
distributions and mean drag force for the model. In
Fig.~\ref{fig:model_dist_f_v_decay}, we show the drag force
distributions for different velocities in (a), and their rescaled form
in (b). Comparison of Fig.~\ref{fig:model_dist_f_v_decay} with the
experimental data in Fig.~\ref{fig:dist_f_v} shows very good
agreement.  Fig.~\ref{fig:model_f_v_decay}a shows the mean drag force
from the simulation, which now has a slow increase with
velocity. Fig.~\ref{fig:model_f_v_decay}b shows the same data on a
log-lin plot.  These results can be fitted by a straight line,
indicating a logarithmically slow increase now built into the
model. This figure compares well with the experimental data in
Fig.~\ref{fig:f_v}. Additionally, this modification to the model does
not qualitatively change the features reported in previous sections.

In summary, the key point of the model is its assumption that the
force chains are modeled as ``springs'' with failure thresholds chosen
from a distribution. Thus, the fluctuations and mean properties of the
drag force are closely associated with the force chain formation and
failure. This understanding is useful in particular because it
underscores the important role of the force chains in granular
systems.  The elastic nature of the model is also interesting, given
the current debate over how forces are transmitted in granular
systems \cite{geng_03b}.

Another interesting observation from the experiments is the seeming
contradiction between the rate-dependence in the mean properties
(e.g., mean drag force vs. velocity, mean avalanche size vs. velocity,
etc.)  and the rate-independence of the fluctuations (e.g.,
rate-independent power spectra, collapse of the avalanche size
distributions, etc.) in the data.  However, this may be understood by
noting that the mean behavior (or the DC part of the signal) is
rate-dependent, while fluctuations (or the AC part of the signal) are
rate-independent. This is also consistent with the failure model we
have discussed; i.e., once the level of the mean behavior is set, the
fluctuating components are subsequently set by the mean behavior.

\section{Conclusions}

To conclude, through experiments and simple failure models, we have
characterized the drag force experienced by an object moving slowly
through a 2D granular material consisting of bidisperse disks. The
drag force is dominated by force chain structures in the bulk of the
system.  The formation and failure of the force chains leads to strong
fluctuations.

We have considered the effect of three control parameters: the medium
velocity, the packing fraction and the tracer particle
size. Experimentally, we find that the mean drag force grows slowly
(logarithmically) with the drag velocity, increases rapidly
(power-law) with the packing fraction above a critical value, and
varies nonlinearly with the size of the tracer particle. The system
exhibits strong statistical invariance in the sense that many physical
quantities collapse into a single curve under appropriate scaling:
force distributions P($f$) collapse when scaled by the mean force,
power spectra P($\omega$) collapse when scaled by the drag velocity,
and avalanche size and duration distributions collapse when scaled by
the mean values of these quantities.

We also show that the system can be understood using a simple failure
model, which reproduces many experimental observations including: a
power law with exponent $\alpha=-2$ for the high-frequency portion of
the power spectrum, exponential distributions for the avalanche size
and duration, and an exponential fall-off at large forces for the
force distributions. The logarithmic increase of the mean force with
the drag velocity can also be accounted for if slow relaxation of the
material is included.

A number of questions remain.  One of these is the nonlinear
dependence of the drag force on the particle diameter.  Heuristically,
one might expect that the drag force would grow linearly in proportion
to the number of force chains contacting the tracer, and that this
would lead to a linear variation of the drag force with diameter.  In
this regard, the fact that the tracers used here were only somewhat
larger than the grains is likely to be important.  Obviously, the
presence of weak rate dependence in the mean force is of interest, and
its origin is still not clear.  The relative elasticity of the
particles (vs. the driving machinery) may be important in this regard,
and future investigations with harder particles would be of interest.
The frictional character of the drag force in the dense regime is
clear in these experiments.  It would be of interest to see what
occurs as the packing fraction is reduced. below $\gamma_c$.  In the
present experiments, the particles experience friction with the base,
so that it is not possible to investigate the gas-like regime.

\acknowledgements

We appreciate helpful interactions with R. Hartley and J. Matthews.
The work was supported by the US National Science Foundation
under Grant DMR-0137119, DMS-0204677, DMS-0244492, and by NASA under
Grant NAG3-2372.

\begin{figure}[h]
\center{\parbox{6in}{\psfig{file=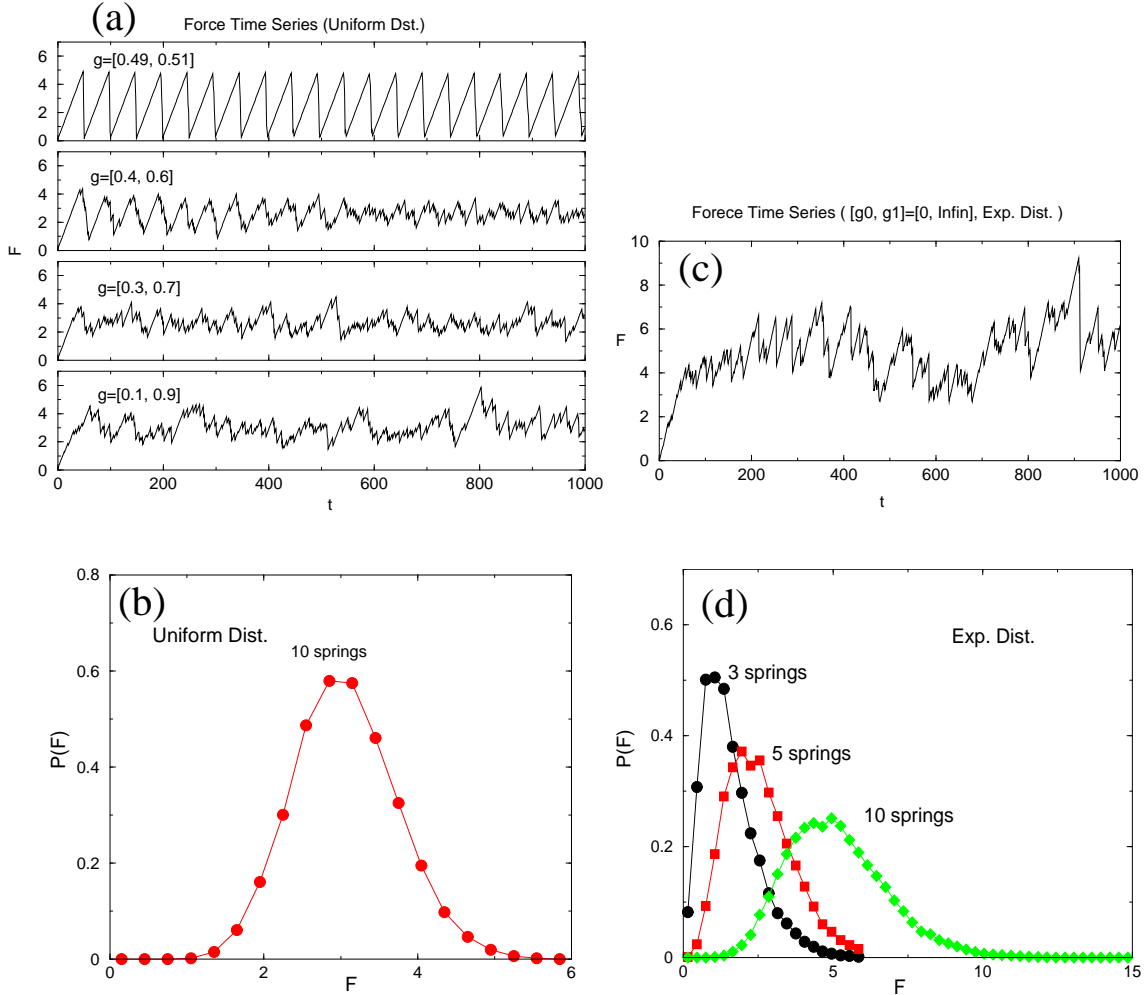,width=6in}}}
\caption{ (a) Force time series generated by the model for various
widths of the threshold band, $[g_0, g_1]$, where the threshold is
uniformly distributed between the $g_0$ and $g_1$. When the threshold
band is narrow, the time series follows a saw-tooth pattern, and when
the threshold band is widened, the force time series become more and
more random. (b) The force distribution of a random force time series
with $[g_0, g_1]=[0.1, 0.9]$, and otherwise as in (a). (c) A random
force time series generated from the model with an exponential
distribution for the threshold. (d) Force distributions derived from
random force time series, such as the one shown in (c), for various
numbers of springs. In contrast to (b), the force distribution with an
exponentially distributed threshold is non-symmetrical, and more
closely resembles the experimental data.}
\label{fig:model_time_series}
\end{figure}

\begin{figure}[h]
\center{\parbox{4in}{\psfig{file=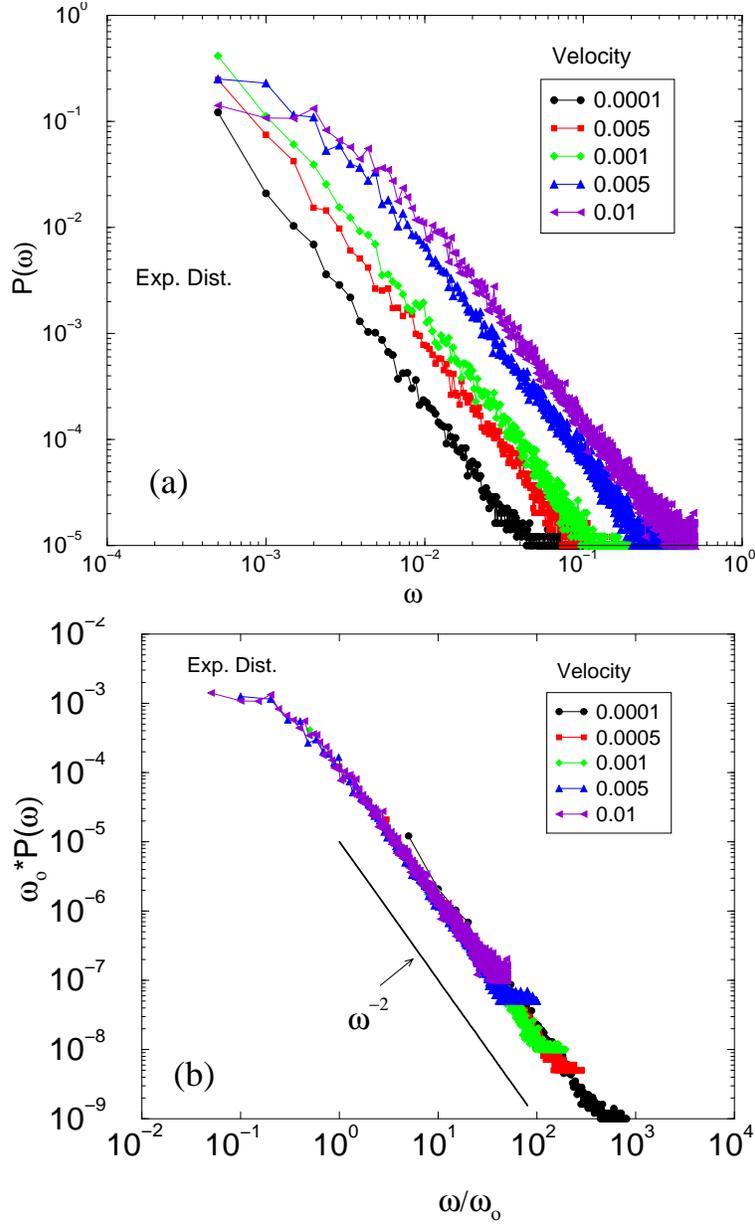,width=4in}}}
\caption{(a) Power spectra, $P(\omega)$, of the force time series at
different rotation rates generated from the model. (b) The scaled
power, $\omega_o P(\omega)$ is plotted against the scaled frequency,
$\omega/\omega_o$, where $\omega_o$ is the rotation rate. The rescaled
power spectra collapse nicely, and follow a power law decay with an
exponent of $-2$ for higher frequencies, in good agreement with the
experimental data shown in Fig.~\protect \ref{fig:power_spec_v}.}
\label{fig:model_power_spec}
\end{figure}

\begin{figure}[h]
\center{\parbox{4in}{\psfig{file=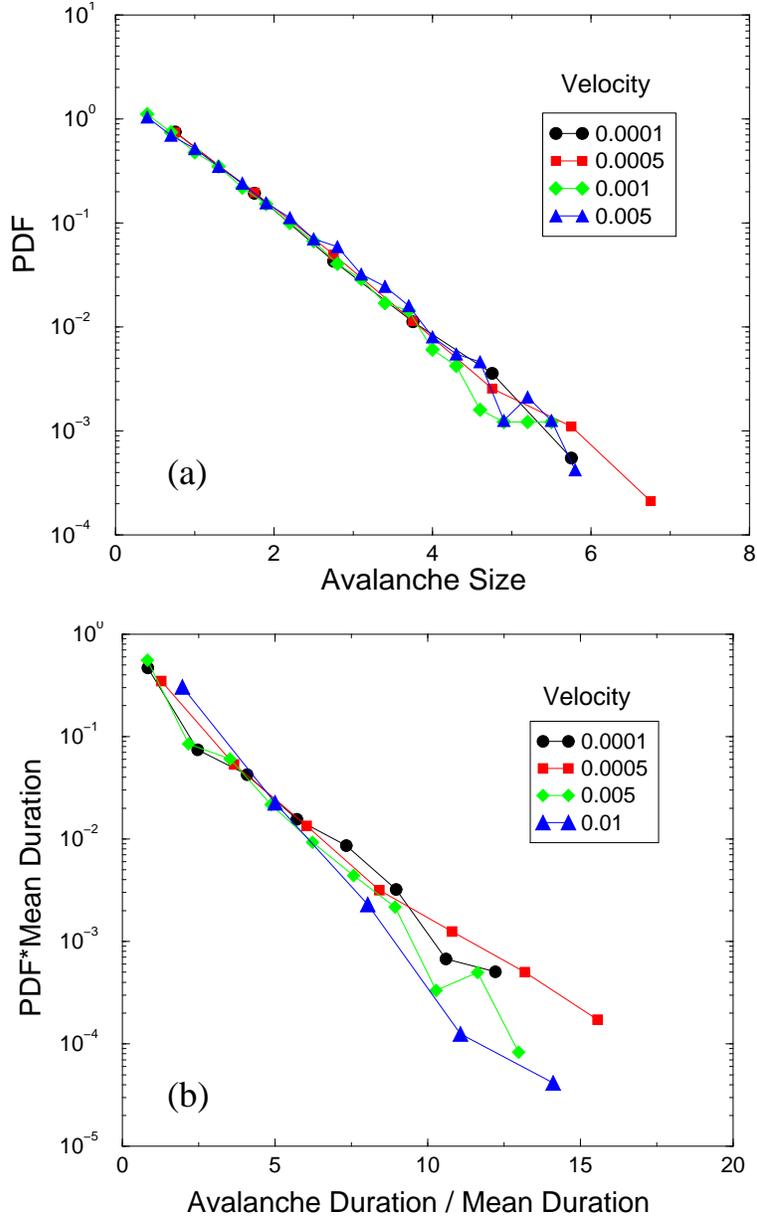,width=4in}}}
\caption{(a) Distributions of avalanche sizes derived from the model.
(b) Rescaled distributions of avalanche durations derived from the
model simulations, where the horizontal axis is divided by the mean
avalanche duration and the vertical axis is multiplied by the mean
avalanche duration. Both distributions of avalanche size and duration
are exponential, in agreement with experimental data shown in
Fig.~\protect \ref{fig:ava_size_dist}.  However, note that the size
distributions in this figure are not scaled while the size
distributions in \protect \ref{fig:ava_size_dist}a are.}
\label{fig:model_ava_size_dist}
\end{figure}

\begin{figure}[h]
\center{\parbox{4in}{\psfig{file=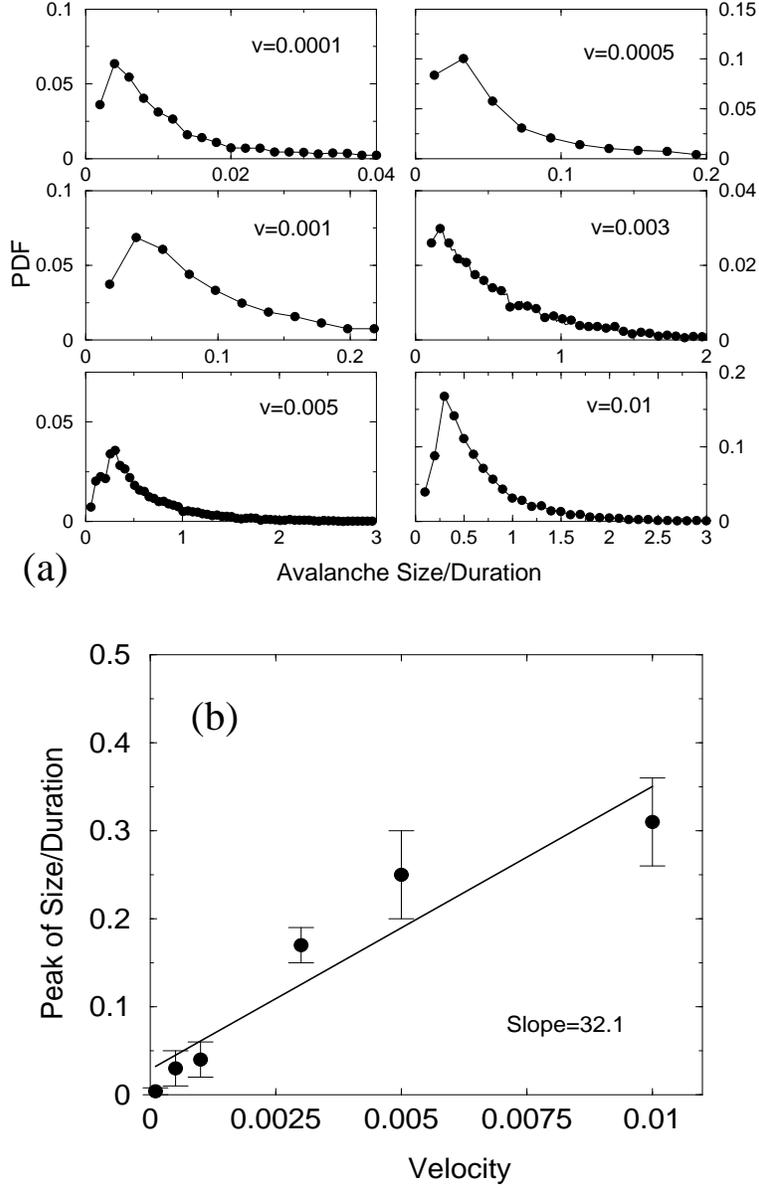,width=4in}}}
\caption{ (a) Distributions of avalanche rates for different rotation
rates derived from the model. (b) The force chain force constant,
$k_{eff}$, obtained by fitting a straight line to a plot of the peak
of avalanche rate vs. the medium velocity. The value $k_{eff}=30.7$
is of the same order of magnitude as $nk_0$, where $n=10$ and
$k_0=1$. This figure compares well with experimental data shown in
Fig. \protect \ref{fig:force_chain_const}. }
\label{fig:model_force_chain_const}
\end{figure}

\begin{figure}[h]
\center{\parbox{4in}{\psfig{file=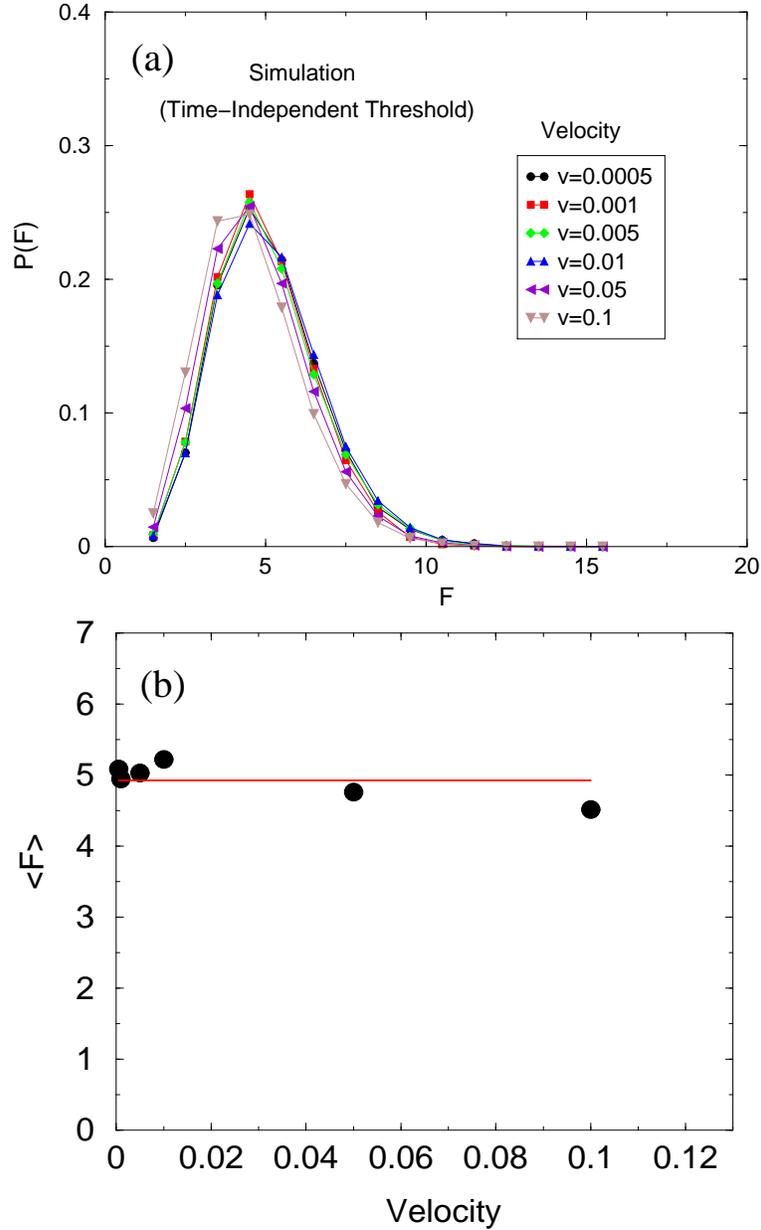,width=4in}}}
\caption{ (a) Model force distributions at different medium
velocities. (b) The mean force vs. the medium velocity from model
simulations. Both (a) and (b) show that the mean force is independent
of the rate. This figure illustrates the problem that the model, so far,
cannot account for the experimental finding that the mean force
increases slowly with the medium velocity.}
\label{fig:model_f_v_no_decay}
\end{figure}

\begin{figure}[h]
\center{\parbox{4in}{\psfig{file=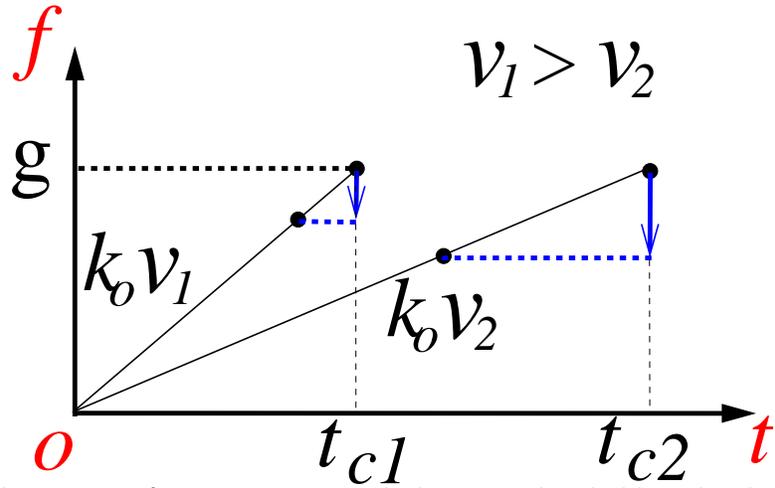,width=4in}}}
\caption{ An illustration of two processes with decaying
thresholds. The slower process ($v_2$ has a waiting time ($t_{c2}$)
and the faster process has a shorter waiting time ($t_{c1}$). The
longer the waiting time, the more the threshold decreases.}
\label{fig:decay_g}
\end{figure}

\begin{figure}[h]
\center{\parbox{4in}{\psfig{file=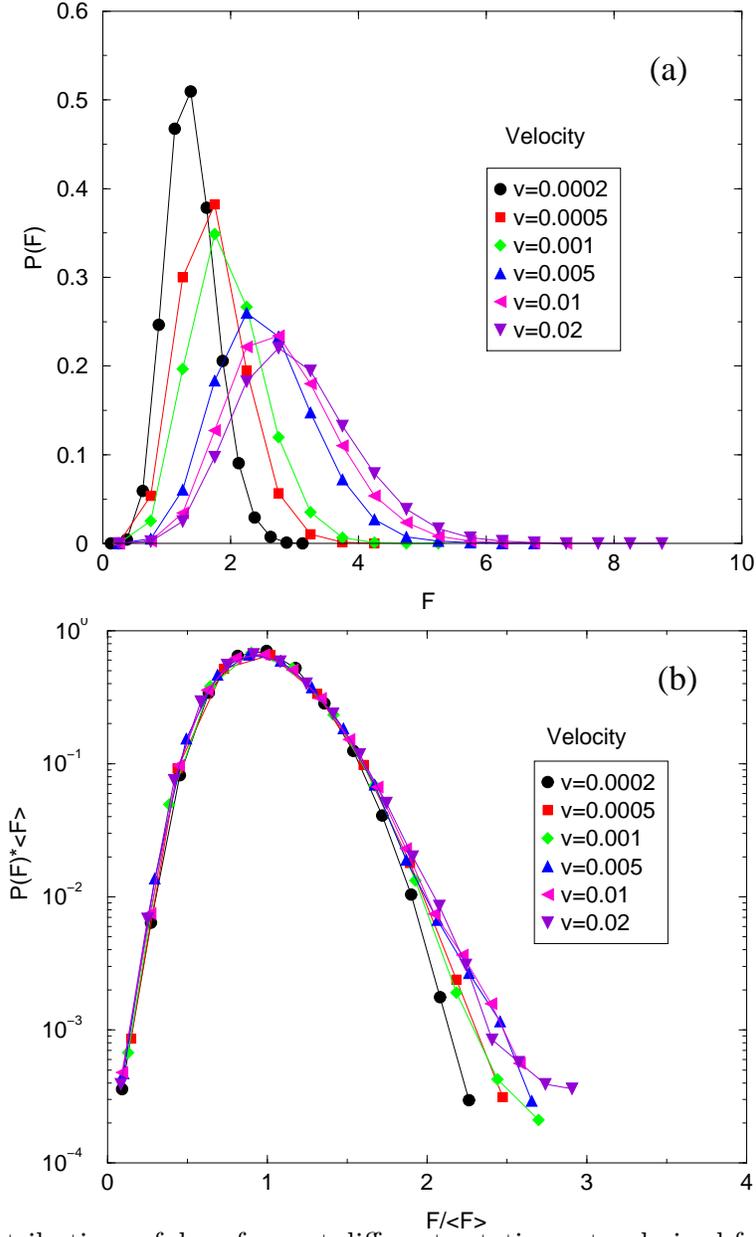,width=4in}}}
\caption{ (a) Distributions of drag force at different rotation rates
derived from the simulations when a decaying threshold is used. (b)
Same data as (a), but with the horizontal axis rescaled by the
corresponding mean force, and the vertical axis multiplied by the mean
force. This figure compares well with experimental data shown in
Fig. \protect \ref{fig:dist_f_v}. }
\label{fig:model_dist_f_v_decay}
\end{figure}

\begin{figure}[h]
\center{\parbox{4in}{\psfig{file=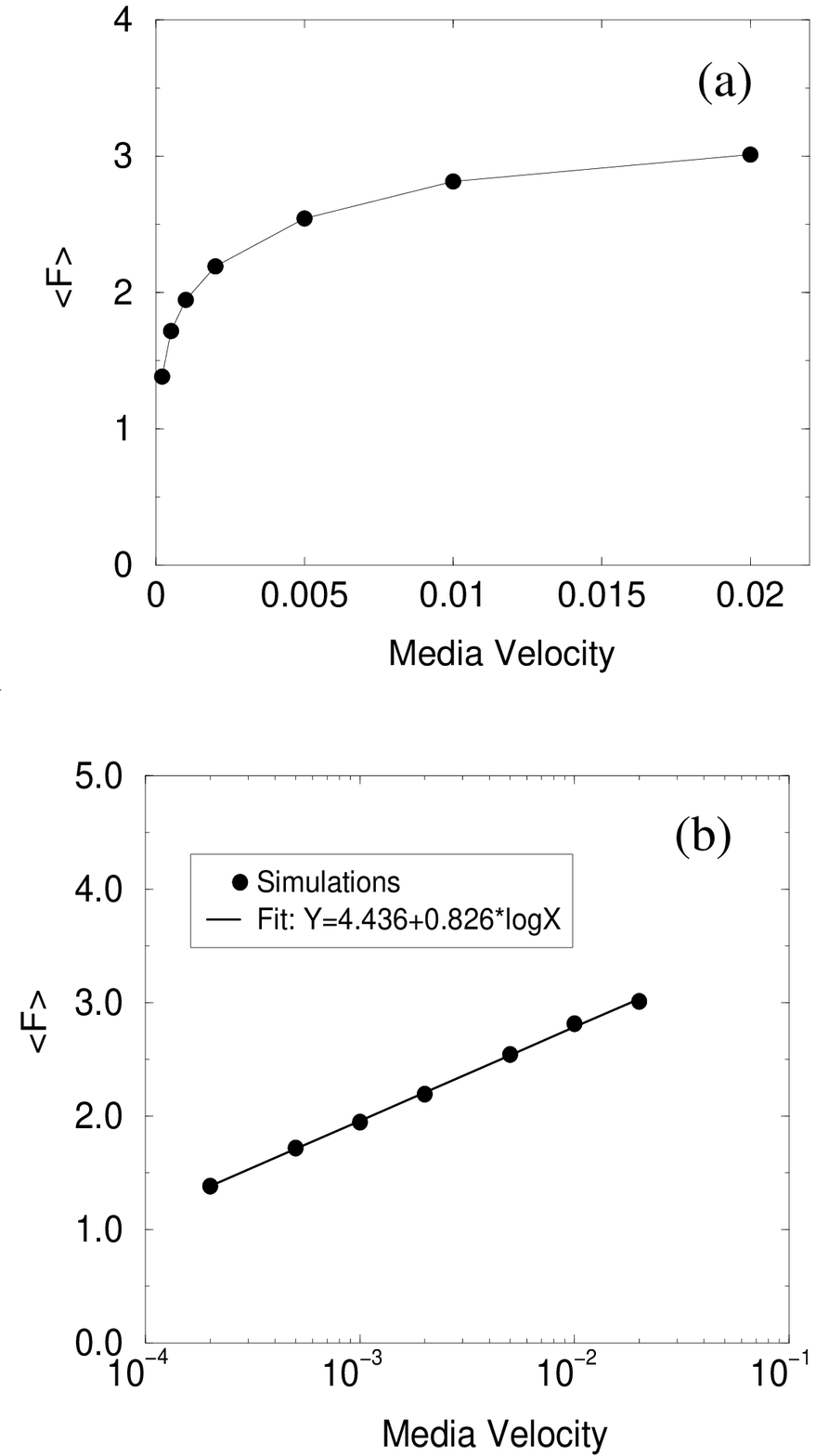,width=4in}}}
\caption{ When slow relaxation of the threshold is considered, the
model is able to account for the slow increase of the mean force as
the medium velocity is increased. (a) The mean drag force, $<F>$, as a
function of medium velocity derived from the simulations. (b) Same
data as (a), but on a log-lin scale to show the logarithmic
increase. This figure compares well with the experimental data of
Fig. \protect \ref{fig:f_v}. }
\label{fig:model_f_v_decay}
\end{figure}

\end{document}